\documentclass{ar2ejp}

\usepackage[dvips]{graphicx,epsfig,color}
\usepackage{amsmath}
\newcommand{\base}{\baselineskip 16pt}
\usepackage{cite}
\usepackage{booktabs}
\usepackage{aas_macros}

\begin{document}

\input psfig.sty

\newcommand{\n}{\ensuremath{\mathrm{n}}}

\renewcommand{\H}{\ensuremath{\mathrm{p}}}

\newcommand{\h}[1]{\ensuremath{{}^{#1}\mathrm{H}}}

\newcommand{\he}[1]{\ensuremath{{}^{#1}\mathrm{He}}}

\newcommand{\li}[1]{\ensuremath{{}^{#1}\mathrm{Li}}}

\newcommand{\be}[1]{\ensuremath{{}^{#1}\mathrm{Be}}}

\newcommand{\Ob}{\ensuremath{\Omega_b}}

\newcommand{\Obh}{\ensuremath{\Omega_bh^2}}

\newcommand{\hef}{\ensuremath{{}^4\mathrm{He}}}

\newcommand{\het}{\ensuremath{{}^3\mathrm{He}}}

\newcommand{\lisx}{\ensuremath{{}^6\mathrm{Li}}}

\newcommand{\lisv}{\ensuremath{{}^7\mathrm{Li}}}

\newcommand{\bes}{\ensuremath{{}^7\mathrm{Be}}}

\newcommand{\beet}{\ensuremath{{}^8\mathrm{Be}}}

\newcommand{\ben}{\ensuremath{{}^9\mathrm{Be}}}

\newcommand{\hes}{\ensuremath{{}^6\mathrm{He}}}

\newcommand{\xm}{\ensuremath{X^-}}

\newcommand{\xp}{\ensuremath{X^+}}

\newcommand{\xz}{\ensuremath{X^0}}

\newcommand{\bex}{\ensuremath{(\bes\xm)}}

\newcommand{\fr}[2]{\ensuremath{\frac{#1}{#2}}}

%new macros

\newcommand{\deut}{\ensuremath{\mathrm{D}}}

\newcommand{\trit}{\ensuremath{\mathrm{T}}}

\newcommand{\CMB}{\ensuremath{\mathrm{CMB}}}

\newcommand{\BBN}{\ensuremath{\mathrm{BBN}}}

\newcommand{\SBBN}{\ensuremath{\mathrm{SBBN}}}

\newcommand{\SM}{\ensuremath{\mathrm{SM}}}

\newcommand{\Orderof}[1]{\ensuremath{\mathcal{O}(#1)}}

%units

\newcommand{\eV}{\ensuremath{\mathrm{eV}}}

\newcommand{\keV}{\ensuremath{\mathrm{keV}}}

\newcommand{\MeV}{\ensuremath{\mathrm{MeV}}}

\newcommand{\GeV}{\ensuremath{\mathrm{GeV}}}

\newcommand{\TeV}{\ensuremath{\mathrm{TeV}}}

%left for copy and paste of older text

\newcommand{\hefm}{\hef}

\newcommand{\hetm}{\het}

\newcommand{\lisxm}{\lisx}

\newcommand{\lisvm}{\lisv}

\newcommand{\besm}{\bes}

\newcommand{\beetm}{\beet}

\newcommand{\benm}{\ben}

\newcommand{\hesm}{\hes}

\newcommand{\bexm}{\bex}

\def\bee{\begin{equation}}

\def\ee{\end{equation}}

\def\epstwo@scaling{0.48}

\def\showtwo#1#2{
\centering
\leavevmode
\epsfxsize=\epstwo@scaling\linewidth
\epsfbox{#1.eps} \hfil
\epsfxsize=\epstwo@scaling\linewidth
\epsfbox{#2.eps}
}

\newcommand{\simge}{\,{}^>_{\sim}\,}
\newcommand{\simle}{\,{}^<_{\sim}\,}

\title{Big Bang Nucleosynthesis as a \\Probe of New Physics}

\markboth{Big Bang Nucleosynthesis as a Probe of New Physics}{Big Bang Nucleosynthesis as a Probe of New Physics}

\author{\\Maxim Pospelov\thanks{pospelov@uvic.ca}
\affiliation{Perimeter Institute for Theoretical Physics, Waterloo,
ON, N2L 2Y5, Canada\\
Department of Physics and Astronomy, University of Victoria, Victoria, BC, V8P
1A1, Canada
}
Josef Pradler\thanks{jpradler@perimeterinstitute.ca}
\affiliation{Perimeter Institute for Theoretical Physics, Waterloo,
ON, N2L 2Y5, Canada
}}

\begin{keywords}
  Big bang nucleosynthesis, early Universe, abundances of light
  elements, extensions of the Standard Model, dark matter
\end{keywords}

\begin{abstract}
  Big bang nucleosynthesis (BBN), an epoch of primordial nuclear
  transformations in the expanding Universe, has left an observable
  imprint in the abundances of light elements. Precision observations
  of such abundances, combined with high-accuracy predictions, provide
  a nontrivial test of the hot big bang and probe non-standard
  cosmological and particle physics scenarios. We give an overview of
  BBN sensitivity to different classes of new physics: new particle or
  field degrees of freedom, time-varying couplings, decaying or
  annihilating massive particles leading to non-thermal processes, and
  catalysis of BBN by charged relics.
\end{abstract}

\maketitle

\base

\section{INTRODUCTION}

The model of the hot expanding Universe, once the subject of fierce
scientific and philosophical debate, is now a well-established
reality. Explosive progress in the field of cosmology over the past
decade allowed for something that may well have been totally
unanticipated in the previous decades: the high-precision
determination of several cosmological parameters. This recent progress
primarily cla\-rifies the state of the Universe at redshifts $z\simle
$few and exposes the relevant physics of the Universe traced back to
the epoch of the decoupling of the cosmic microwave background (CMB)
at $z\simle 10^3$.  At the same time, these recent advances also allow
one to check for the consistency with one the building blocks of
modern cosmology: big bang nucleosynthesis (BBN).  The agreement, or
rather the lack of qualitative disagreement between cosmology in the
$z = 0-10^3$ interval and BBN at $z \sim 10^9$, is very important, as
it leads to the conclusion that the very early Universe was governed
by the same physical laws of nature as the current Universe and that
it contained very similar if not identical particle content.
Remarkably, that the transformations and the synthesis of light
elements in the expanding Universe occur with the active participation
of all the interactions known to date: strong, weak, electromagnetic,
and gravitational.  It is well known that even a mild modification of
the standard conditions in the early Universe, at the time of BBN or
in the subsequent evolution, may lead to observable deviations in
primordial abundances.  Thus, employing the precise determinations of
the primordial abundances allows one to set limits or constrain
various scenarios with deviations from General Relativity or from
particle physics of the Standard Model (SM).

Primordial nucleosynthesis is not the earliest cosmological epoch.  It
must have been preceded by an era in which a mechanism
capable of producing the observed baryon-antibaryon asymmetry of the
Universe  (baryogenesis) was operative. Moreover, a period of generation of nearly
scale-invariant cosmological perturbations---for which the leading
candidate is inflation---had set the initial conditions from which the
large-scale structure of the Universe evolved.
Neither the cosmological timing nor the main ``players'' in inflation
or baryogenesis---such as the inflaton, the right-handed neutrino, and
so on---are known with certainty. Although many successful models of
this earliest epoch(s) exist, it is extremely difficult to directly
test the associated physics in the laboratory. Thus, a multitude of
baryogenesis- and inflationary scenarios may be viable.  In contrast,
BBN relies on very well-studied pieces of physics, such as spectra and
reactions of light elements and their weak decays. It involves almost
no free parameters and occurs in a well-understood sequence of events
in the early Universe.  The consistency of BBN predictions with
observations imposes an important calibration point on all new models
of particle physics that require them to achieve some degree of
``normalcy'' before $t=1$~s.  This requirement places powerful
constraints on many extensions of the SM.

The purpose of this review is to reveal different physical mechanisms
by which new physics can affect BBN and its light element abundance
predictions, which have been extensively discussed in the literature
for as long as the BBN theory has existed. In fact, one of the first
papers on the subject by Alpher, Follin and Herman \cite{AFH} points
out that the nature of the neutrino species (Dirac versus Majorana)
affects the neutron-to-proton freeze-out ratio, $n/p$, and thereby the
primordial helium abundance. Since then, a great deal of research
dedicated to non-standard BBN scenarios has been performed, and a
number of different ways in which new physics can change the outcome
of BBN have been identified. Here, we present a non-technical review
of many aspects related to these interesting possibilities.

The developments of the past decade also confirmed the conclusion that
the SM is not the ultimate theory. Important pieces of the puzzle come
from cosmological observations suggesting that the energy balance of
the modern Universe is dominated by dark energy and dark matter.  Is
it possible that physics related to these mysterious substances
interfered with the sequence of events that ultimately led to the
fusion of the primordial elements?
Although there is no compelling evidence that this occurred, some new
physics models related to particle dark matter may reduce the tension
between data and the standard BBN theory (SBBN) prediction for the
lithium isotopes. Even though the resolution of this tension may be
related to astrophysics, it is nevertheless intriguing to entertain
the possibility that the relics of the early Universe may hint at
deviations from the SM.  Future developments in astrophysics,
cosmology, and particle physics may help to clarify this question.

\subsection{Universe at redshift of a billion: basic assumptions and
  main stages of BBN}

The primary activity of BBN took place in the era associated with
photon temperatures between $T\simeq {\rm few}\,$MeV and $T\simeq
10\,$keV, in the cosmic time window $t\simeq (0.1\div10^4)\,$s, and
may be considered as a transition from a neutron-proton statistical
equilibrium with no other nuclear species to a Universe with a
significant presence of helium. BBN produced the bulk of \he4 and D as
well as good fractions of \he3 and \li7 observed in the current
Universe. All the other elements are believed to be produced either by
stars or by cosmic rays.

We begin by specifying the main assumptions on which SBBN theory
rests:

\begin{enumerate}

\item It is assumed that the Universe is spatially homogeneous and
  isotropic and that its energy density is completely dominated by
  radiation so that physical distances scale as $a(t) \propto
  t^{1/2}$.  It is also assumed that the space-time geometry is flat
  with zero spatial curvature.  The theory of general relativity
  relates the expansion rate with the time elapsed since the big bang
  and with the energy density of the ambient cosmological fluid
  $\rho$:
  \begin{equation} 
  H \equiv \frac{\dot a}{a} =
   \sqrt{8\pi G_N \rho/3} \simeq \frac{1}{2t} ,
   \label{Hubble1} 
 \end{equation} 
 where the energy density of the radiation-dominated Universe scales
 as $\rho \sim a^{-4} \sim T^4$; $G_N$ denotes Newton's constant $G_N
 = (8 \pi M_{\rm Pl}^2)^{-1}$ where $M_{\rm Pl}\simeq 2.43\times
 10^{18}\,$GeV is the reduced Planck mass. It is also assumed that the
 dark matter and dark energy components of the Universe are
 ``well-behaved'', that is, that their contribution to the energy density
 at the time of BBN is negligible.

\item It is assumed that the initial temperature of the
  radiation-dominated epoch of the Universe was well above the
  neutron-proton mass difference $\Delta m_{np} = 1.293$~MeV, so that
  the initial conditions for the nuclear reaction framework are well
  specified:
  \begin{equation} \left. (n_n \simeq n_p) \right|_{T\gg \Delta
      m_{np}} = \fr12 n_b.  \end{equation} The energy-momentum
  distribution of neutrons and protons is very close to being thermal.
  Moreover, a near-perfect spatial homogeneity for the
  distributions of neutrons and protons is assumed.

\item It is assumed that the particle content and their interactions
  are given by that of the SM and that by the time of BBN the baryon
  asymmetry was already present. The energy density is completely
  dominated by photons and the three species of SM neutrinos (as well
  as electrons and positrons before their annihilation), whereas
  neutrons and protons carry only a negligible fraction of the total
  energy density.  The standard field content implies that there was
  an ``uneventful'' cosmological period between the time of BBN and
  the epoch of hydrogen recombination with electrons which implies
  that the baryon-to-entropy density ratio measured at the period of
  CMB-``formation'' directly translates into the one at BBN:
  \begin{equation} 
    \frac{n_b}{s}(t_{\BBN}) = \frac{n_b}{s}(t_{\CMB}).
  \end{equation}
  Indeed, the measurements of the Wilkinson Microwave Anisotropy Probe
  (WMAP) satellite have allowed us to pinpoint the baryon density for
  a standard $\Lambda$CDM cosmology, that is a flat Universe filled
  with baryons, cold dark matter, neutrinos and a cosmological
  constant, to an accuracy of better than
  $3\%$~\cite{Dunkley:2008ie}. Expressed in terms of the
  baryon-to-photon ratio,
  \begin{equation} 
    \eta_b (t_{\CMB}) =
    \frac{n_b}{n_\gamma}(t_{\CMB}) = (6.23 \pm 0.17)\times
    10^{-10} \label{etaCMB} ,
  \end{equation} 
  this number provides a measure of the nucleon content of the
  Universe at BBN.

\item Finally, it is assumed that the properties of particles and
  nuclei (masses, couplings, scattering cross sections and lifetimes)
  are identical between their current values and their values at
  $t_{\BBN}$.  This condition, in fact, follows from the assumption of
  a minimal field content, as hypothetical changing couplings and
  masses would necessitate new ultra-soft scalar fields.

\end{enumerate}

In many theories with non-standard cosmological or particle physics
content, some of these assumptions can be violated. For example,
``late decays'', that is, decays of metastable heavy particles during
or after BBN can lead to energy injection into the primordial plasma
and thus to temporary but strong departures from thermal equilibrium
for some species. The presence of the new relativistic degrees of
freedom may affect the total energy density of the Universe, and, by
modifying the Hubble expansion rate, may change the outcome of nuclear
reactions. Before we consider such modifications in more detail, we
remind the reader of the main components of the SBBN and briefly
review its current observational status.

SBBN theory, including the hits and misses of the original papers
\cite{ABG,AH,Hayashi}, is now well understood and described in detail
in many text-books and reviews
\cite{Malaney:1993ah,Sarkar:1995dd,Iocco:2008va,Steigman:2007xt,Jedamzik:2009uy}.
SBBN comprises the set of first-order differential Boltzmann equations
on the abundances of the different species,
\begin{eqnarray}
\label{start}
\frac{dY_i}{dt}=-H(T)T \frac{d Y_i}{ dT } = \sum (\Gamma_{ij}Y_j+ \Gamma_{ikl}  Y_k Y_l+...),
\end{eqnarray}
where $Y_i = n_i / n_b$ are the time $t$ (or temperature
$T$)-dependent ratios between the number density $n_i$ and the baryon
number density $n_b$ of light elements $i=\,p$, $n$, D,\he4, and so
on. The $\Gamma_{ij...}$ represent generalized rates for element
interconversion and decay that can be determined in experiments and/or
inferred from theoretical calculations.  $H(T)$ is the
temperature-dependent Hubble expansion rate from Eq.~(\ref{Hubble1}).

The full form of the (non-integrated) Boltzmann equations should be
given in terms of particle distribution functions over energy and
momenta. However, in practice, the system of equations~(\ref{start}),
which assumes thermal distributions for nuclei, provides an excellent
approximation because the frequent interactions with the numerous
$\gamma$s and $e^{\pm}$s in the plasma keep the light elements tightly
coupled to the radiation field. The dependence of $H(T)$ on the
temperature of the primordial plasma can be further specified:
\begin{equation}
\label{Hubble2}
H(T) = T^2 \left(\fr{8\pi^3 g_* G_N}{90}\right)^{1/2} ,~~ {\rm where}
~ g_* = g_{boson} +\fr78g_{fermion}\, , 
\end{equation} 
where the $g$s denote the excited relativistic degrees of freedom.
This expression needs to be interpolated across the electron-positron
annihilation epoch, in which the photon bath is heated with respect to
the neutrino reservoir. The neutrinos maintain a quasi-thermal
spectrum with temperature
\begin{equation} 
T_\nu \simeq (4/11)^{1/3} T,
\label{Tnu}
\end{equation} 
in the approximation of full neutrino decoupling at the time of
electron-positron annihilation [with small calculable corrections
\cite{Dolgov:1992qg}].  Following $e^\pm$ annihilation, the Hubble rate
is given by $H(T) \simeq T_9^2/(2\times 178~{\rm s})$, where $T_9$
denotes the photon temperature $T$ in units of $10^9\, \mathrm{K}$.

\begin{figure}[t!]
\centerline{\includegraphics[width=\textwidth]{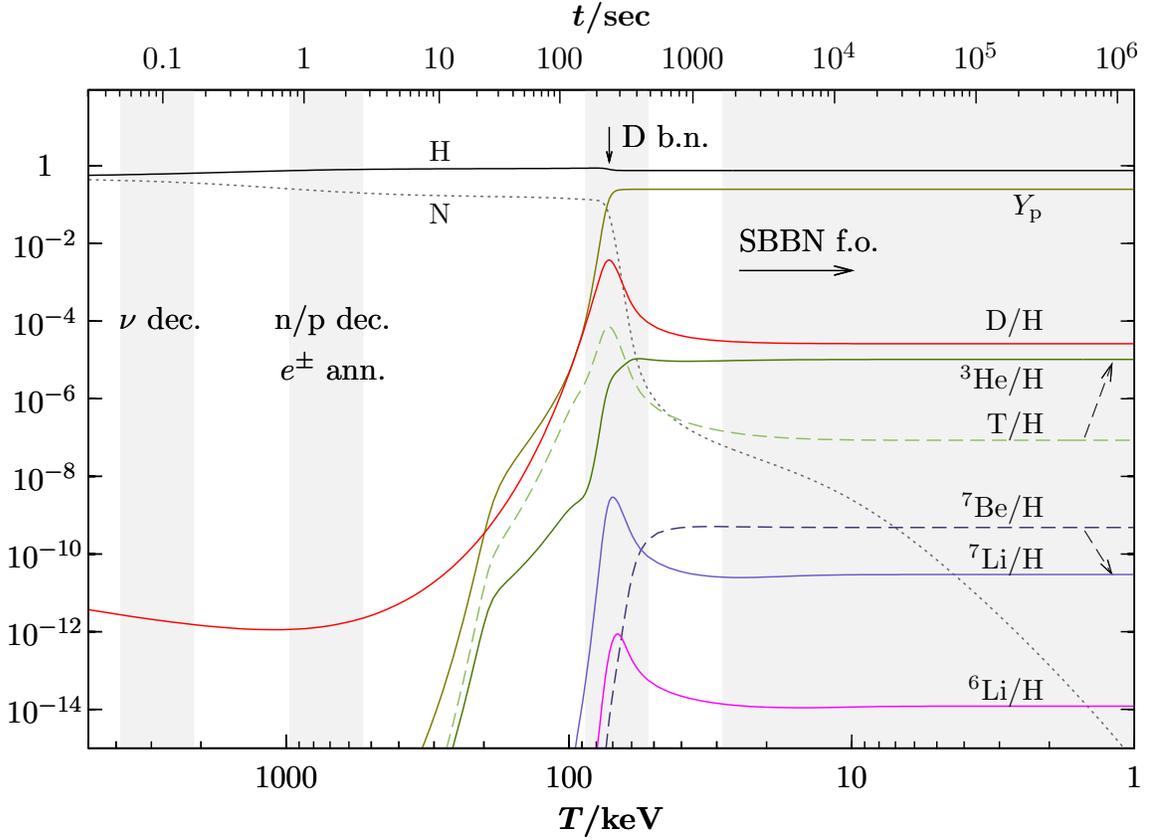}}
\caption{\base Time and temperature evolution of all standard big bang
  nucleosynthesis (SBBN)-relevant nuclear abundances.  The vertical
  arrow indicates the moment at $T_9 \simeq 0.85$ at which most of the
  helium nuclei are synthesized.  The gray vertical bands indicate
  main BBN stages. From left to right: neutrino decoupling,
  electron-positron annihilation and $n/p$ freeze-out, D bottleneck,
  and freeze-out of all nuclear reactions. Protons (H) and
  neutrons (N) are given relative to $n_b$ whereas $Y_p$ denotes the
  \hef\ mass fraction.}
\label{figure1}
\end{figure}

A number of well-developed integration codes, pioneered by Wagoner et
al.~\cite{Wagoner:1966pv}, allow one to solve the BBN system of
equations numerically and to obtain the freeze-out values of the light
elements. Also, good semi-analytic results can be obtained, see {\it
  e.g.} Ref.~\cite{Mukhanov:2003xs}.  The code that we use for this
review is based on that of Ref.~\cite{Kawano:1992ua}. Here we
incorporate some significant improvements and updates; physical
constants, isotope masses, and conversion factors are determined from
the evaluations~\cite{Audi:2002rp,Mohr:2005kk}. For all important SBBN
reactions (\textit{i.e.} up to A=7) we employ the results
of~\cite{2004ADNDT..88..203D} with the exceptions of the
$n(p,\gamma)\deut$ and $\het(\alpha,\gamma)\bes$ reactions for which
we follow~\cite{Ando:2005cz} and \cite{Cyburt:2008up},
respectively. To arrive at an accurate prediction of \hef, we
numerically integrate all weak rates at each time step for which zero
temperature radiative corrections as well as relativistic Coulomb
corrections (when applicable) are taken into
account~\cite{Lopez:1998vk,Wilkinson:1982hu}. Finally, as assessed in
\cite{Lopez:1998vk}, we apply a slight upward-shift of $0.72\%$ of the
resulting \hef\ abundance to account for more subtle, subleading
corrections to the BBN reaction network. We then find very good
agreement between our light element-abundance predictions and those
presented in~\cite{Cyburt:2008kw} at the WMAP value~(\ref{etaCMB}) and
with a neutron lifetime of $\tau_n=885.7\,$s.

Figure~\ref{figure1} shows the output of the SBBN reaction network.
The evolution of all the main isotopes with $A\leq 7$ is plotted as a
function of time and temperature. Beginning from the earliest times,
the following sequence of events occurs: (a) the chemical decoupling
of neutrinos from the thermal bath, (b) the annihilation of electrons
and positrons, (c) the freeze-out of neutrons and protons, (d) an
``intermission'' between the $n/p$ freeze-out and the deuteron
ignition at the end of the ``bottleneck,'' (e) helium synthesis at
$T_9 \simeq 0.85$ ($70\,$keV), and finally (f) a follow-up stage in
which the main nuclear reactions gradually drop out of equilibrium and
the abundances of all light elements freeze out. The freeze-out
abundances are given by the horizontal lines on the right-hand side of
the graph.  Although some neutrons are still generated by residual
deuterium (D) fusion below $T_9 \sim 0.1$ ($10\,$keV), they are too
few in number to cause any further change in the elemental abundances.
Below we discuss the fusion of the light elements and compare their
SBBN predictions with observations.

\subsubsection {\Orderof{0.1} abundances: \hef.}  
The beauty of the SBBN prediction for \he4\ lies in its simplicity.
Only a few factors that determine it. The rates for weak scattering
processes that inter-convert $n\leftrightarrow p$ at high plasma
temperatures scale as $G_F^2 T^5$, where $G_F$ is the Fermi
constant. As the Universe cools, these rates drop below the
$T^2$-proportional Hubble rate $H(T)$ Eq.~(\ref{Hubble2}). The
neutron-to-proton transitions slow down, and the ratio of their
respective number densities cannot follow its chemical-equilibrium
exponential dependence, $n/p|_{\mathrm{eq}} \simeq \exp(-\Delta
m_{np}/T)$. Around $T\simeq 0.7\,\MeV$ this dependence freezes out to
$n/p \simeq 1/6$ but continues to decrease slowly due to residual
scattering and $\beta$-decays of neutrons. The formation of D during
this intermission period is delayed by its photo-dissociation process
that occurs efficiently because of the overwhelmingly large number of
photons [see Eq.~(\ref{etaCMB})] with energies in excess of the
deuteron binding energy $E_d = 2.22\,\MeV$. Once the temperature drops
to $T_9\simeq 0.85$, the exponential Boltzmann suppression of such
photons is sufficient to build a number density in D that is large
enough to ignite other nuclear reactions.  At these temperatures, the
neutron-to-proton ratio has dropped to approximately 1/7 and very
quickly, all neutrons are consumed and are incorporated into \he4
nuclei that have the highest binding energy per nucleon among all
isotopes lighter than carbon. Thus, to a rather good accuracy,
\begin{equation}
  Y_p \simeq \left. \frac{2n/p}{1+n/p} \right|_{T_9 \simeq 0.85 } .
\end{equation}
The \he4 mass fraction $Y_p$ is very weakly dependent on $\eta_b$ as
well as on the precise values for almost all nuclear reaction rates.
Instead, $Y_p$ is sensitive on the {\it timing} of major BBN events,
such as the neutron-to-proton freeze-out and the end-point of the D
bottleneck.  Consequently, the prediction for $Y_p$ relies on such
well-measured quantities as the Newton constant, the neutron-proton
mass difference, the Fermi constant, the neutron lifetime and the
deuteron binding energy.  The sensitivity of \he4 to all possible
non-standard scenarios that modify the timing of BBN events causes
this isotope to serve as the BBN ``chronometer''~\cite{Steigman:2007xt}.
SBBN predicts $Y_p$ with an impressive precision (much better than
$1\%$) so that the error bar is dominated by the uncertainty in the
neutron lifetime.

The determination of primordial \he4 has been carried out by several
groups~\cite{Peimbert:2007vm,Izotov:2007ed}.  Such determination is
performed via observations of hydrogen- and helium-emission lines in
extragalactic HII-regions of low metallicity. Although the statistical
error on $Y_p$---due to a large number of such observations---can be
driven down to a \Orderof{10^{-3}}-level, the extraction of the
primordial value is limited by systematic
errors~\cite{Olive:2004kq}. Indeed, the latter errors are much larger
than both, the statistical error and the accuracy of the SBBN
prediction.  We quote the most conservative analysis by Olive~\&
Skillman~\cite{Olive:2004kq}, together with a recent update on the
SBBN prediction~\cite{Cyburt:2008kw},
\begin{eqnarray}
{\rm SBBN:}&& ~~~Y_p = 0.2486 \pm 0.0002 \\
{\rm extrapolation ~to~ primordial~ value:}&& ~~~Y_p= 0.249\pm 0.009.
\end{eqnarray}
Another recent analysis~\cite{Fukugita:2006} finds a somewhat smaller
error bar, $Y_p\simeq 0.250\pm 0.004$, and for the purpose of this
review we assume that the observational range for $Y_p$ is limited to
the range between 0.24 and 0.26. Clearly this error range is one of
the most important issues in the application of BBN to particle
physics, as it translates directly into the tightness of constraints
on many non-standard scenarios. Further progress in understanding the
error budget in the extraction of the primordial value of $Y_p$ is
needed.

\subsubsection{ \Orderof{10^{-5}} abundances: \deut\ and \het.}

Rapid formation of D at $T_9 \sim 1$ is counter-balanced by the uptake
of helium producing reactions. Below $T_9 \sim 0.8$ the neutron-supply
for D quickly drops in abundance, producing a characteristic peak in
the D/H ratio as a function of temperature, Figure~\ref{figure1}.  The
near complete burning of D then results in a rather small freeze-out
value on the order of few$\times 10^{-5}$. Likewise, \het\ has a
similar abundance to D.  BBN predictions for both elements are very
sensitive to nuclear reaction rates as well as to the $\eta_b$-input.
Nuclear reaction rates relevant for the formation of \het\ and D have
been well measured to better than 10\% accuracy. Therefore, by use of
the current WMAP input, SBBN can make  a fairly precise
prediction for the abundances of these elements~\cite{Cyburt:2008kw}:
\begin{eqnarray}
{\rm SBBN:}&&\quad{\rm D/H} = 2.49\pm 0.17\times 10^{-5}\\
{\rm SBBN:}&&\quad\het/{\rm H } =(1.00\pm 0.07)\times 10^{-5}. 
\end{eqnarray}

D can be observed in the local Universe as well as in highly
redshifted clouds at low metallicities. The latter, see
Refs. \cite{Burles:1997ez,Levshakov:2001xi,Crighton:2004aj,O'Meara:2006mj,Pettini:2008mq},
is the preferred way of determining its primordial fraction.  The
observations of D absorption lines in quasar absorption systems (QALS)
is challenging given that only a tiny isotopic shift separates the D
line from the main hydrogen line. Only in a few cases of QALS with
sufficiently simple velocity structure has the D/H ratio been
extracted.  The scatter of those resulting D/H determinations poses a
certain problem because it exceeds the naively averaged error bar.
Artificially inflating the errors in order to account for this scatter
produces the following result~\cite{Pettini:2008mq}: \begin{equation}
  {\rm QALS~ observations:} \quad \frac{\rm D}{\rm H} = (2.82 \pm
  0.21) \times 10^{-5}.
\label{Dexp} 
\end{equation} This range agrees remarkably well with the SBBN prediction using
the WMAP $\eta_b$-input. Despite the agreement one should be
apprehensive of two potential problems. First, the galactic evolution
of deuterium presents astronomers with a number of puzzles and
implies some significant degree of absorption of D onto dust grains
(For recent discussions, see for instance
\cite{Linsky:2006mr,Prodanovic:2009es}).  Do similar mechanisms exist
at QALS, and if so, should they result in an upward correction to
Eq.~(\ref{Dexp}) that would reflect the true primordial value of D/H?
Second, the origin of scatter in the QALS data is not explained and
could be a sign of underestimated systematic errors {\it or}
additional depletion mechanisms.  Until these issues are better
understood, primordial values as high as $4\times 10^{-5}$ cannot be
convincingly ruled out.

Unlike D, which can be depleted only in the course of the galactic
evolution, both production and depletion of \he3 may occur; therefore,
a direct interpretation of \he3/H measurements as primordial is not
possible.  However, given the uni-directional evolution of D, one can
conclude that the \he3/D ratio will only grow thereby providing us
with a very important {\it upper} bound on the primordial \he3/D
ratio~\cite{Sigl:1995kk}. For the purpose of this review, the
primordial ratio can be constrained
as~\cite{Geiss:2007} \begin{equation} {\rm
    observational~bound:}\quad\frac{\het}{\rm D} < 1, \end{equation}
which does not challenge SBBN, but rather represents an important
constraint on non-standard BBN scenarios.

\subsubsection{ \Orderof{10^{-10}} abundances: {\rm \lisv}.} 
The formation of $A=6,7$ nuclei is suppressed in SBBN due to the
absence of stable $A=5$ elements.  Tritium-$\alpha$ and \he3-$\alpha$
fusion generates \lisv\ and \bes\ nuclei, but with rates that are much
smaller than the Hubble expansion rate.  Consequently, only a tiny
number, few$\times 10^{-10}$ of \lisv\ and \bes\ nuclei relative to
hydrogen is generated. At a later stage of cosmological evolution
\bes\ is ultimately converted into \lisv\ via electron capture.  At
the WMAP-measured value of the baryon-to-photon ratio $\eta_b$, more
than 90\% of the primordial lithium is produced in the form of \bes\
in the radiative capture process, \hef+\het$\to$\bes+$\gamma$. The
output of \bes\ is almost linearly dependent on the corresponding
$S$-factor for this reaction, which was recently re-measured by
several groups \cite{NaraSingh:2004vj,Gyurky:2007qq,Brown:2007sj}.
The current $\sim$15\% accuracy prediction for the combined
\bes+\lisv\ abundance stands at~\cite{Cyburt:2008kw}
\begin{equation}
\label{LiSBBN} {\rm SBBN}:\quad\fr{\lisv}{\rm H}=
5.24^{+0.71}_{-0.67}\times 10^{-10}.  
\end{equation} 
The discrepancy between this prediction and observations is often
referred to as the lithium problem.

Due to its very low abundance, lithium cannot be determined from
observations of extragalactic absorption clouds.  Instead, all
observations of the \li7/H ratio must be performed in the atmospheres
of low-metallicity galactic halo stars. The near constancy of \li7/H
ratios at a low- but wide range of metallicities and for some range of
effective stellar temperature is called the ``Spite plateau''. For a
long time this plateau was thought to be an accurate representation of
the primordial lithium abundance.  A rather small scatter along the
Spite plateau supports the interpretation of the measured \li7 values
as primordial.  Although there have been numerous determinations of
the \li7 abundance on the Spite plateau, the most recent observations
seem to indicate some metallicity-dependence of the \lisv\
abundance. Furthermore, scatter may not be negligible, but rather may
favor some amount of stellar \lisv\ depletion
\cite{Melendez:2009fm,Aoki:2009ce}. Currently, the observational
status of the primordial lithium abundance is given by
\cite{Ryan:1999jq,Bonifacio:2002yx}
\begin{eqnarray}
{\rm Spite~plateau~value:}&&~~\fr{\lisv}{\rm H} = 1.23^{+0.34}_{-0.16}\times 10^{-10},
\label{Spite}
\end{eqnarray}
which is a factor of three to five {\it lower} than the SBBN-predicted
amount of primordial lithium, Eq.~(\ref{LiSBBN}).  It is important to
keep in mind that measurements of lithium in globular clusters have
resulted in somewhat higher abundances of $(2.19\pm 0.28)\times
10^{-10}$ \cite{Bonifacio:2002yx} (for other observational
determinations of the \lisv\ abundance consistent with
\cite{Ryan:1999jq,Bonifacio:2002yx}
see~\cite{Pasquini:1996tp,Asplund:2005yt}). The explanation of the
discrepancy between (\ref{LiSBBN}) and (\ref{Spite}) could be of
purely astrophysical origin.  [A potential additional depletion of
SBBN values would require some un-orthodox modifications of secondary
\bes-destroying reactions~\cite{Coc:2003ce,Cyburt:2009cf}---but so far
this possibility has not found support from nuclear
experiments~\cite{Angulo:2005mi}].  Atmospheric \li7 may have been
partially depleted from atmospheres of population II stars due to
additional settling mechanisms. Also, it has been suggested that
stellar models that assume a factor $\approx 2$ suppression of \lisv\
in such stars, are consistent with observations of other
abundances~\cite{Richard:2004pj,Korn:2006tv}.  Finally, it is also
conceivable that the lithium problem points directly towards physics
beyond the SBBN model, perhaps related to new physical processes at
$T_9 \sim 0.5$. Unfortunately, given the controversial status of
\lisv\, it is difficult to use this isotope for constraining models of
new physics that modify its abundance. Instead, we point to certain
interesting new physics possibilities that might be responsible for
the reduction of \lisv/H.

\subsubsection{ \Orderof{10^{-14}} and less abundances: \lisx\ and $A\ge 9$ elements.}
\lisx~is the only stable $A=6$ element. The main SBBN reaction that
produces it,
\begin{equation}
\hef + {\rm D} \to \lisx +\gamma,\quad Q= 1.47 {\rm MeV},
\label{He4DLi6}
\end{equation}
is suppressed by four orders of magnitude relative to other radiative
capture reactions such as $\hef+\trit\to\lisv+\gamma$, and is
suppressed by approximately seven to eight orders of magnitude
relative to other photonless nuclear rates. The reason for the extra
suppression is unique as it arises from almost the same charge-to-mass
ratio for \hef\ and D, which inhibits the E1 transition, thereby
making this radiative capture extremely inefficient. At the same time,
the proton reaction that destroys \lisx\ is very fast which leads to a
$\Orderof{10^{-14}}$-level prediction for primordial \lisx.  This is
well below modern detection capabilities.  Heavier elements with $A\ge
9$ such as \ben, $^{10}$B and $^{11}$B are never produced in
significant quantities in the SBBN framework because of the absence of
stable $A=8$ nuclei, as \beet\ is under-bound by 92~keV and decays to
two $\alpha$ particles.

Detections of \ben, $^{10}$B, and $^{11}$B have been made for many
stars at low metallicities and are not controversial. Observations of
\ben\ \cite{Primas:2000gc,Boesgaard:2005pf} are far above the
$\Orderof{10^{-18}}$ SBBN prediction and exhibit a linear correlation
with oxygen, clearly indicating its secondary (spallation) origin
\cite{Fields:2004ug}. The lowest level of detected \ben/H is at
$\sim\,$few$\times 10^{-14}$, which translates into a limit on the
primordial fraction of $2\times 10^{-13}$~\cite{Pospelov:2008ta},
assuming that there is no significant depletion of \ben\ in stellar
atmospheres.

Because SBBN predicts very low \li6 abundances, it is usually assumed
that the origin of this isotope is not associated with BBN.
Conversely, any level of \li6 detection in stars at low metallicity in
excess of $\Orderof{10^{-12}}$ is of interest for the test of both the
SBBN framework and the galactic chemical evolution models.
Remarkably, the existence of a \li6 plateau at an isotopic ratio of
$\li6/\li7\simeq 0.05$ has been claimed from ($2\sigma$) detections in
approximately $\sim 10$ low-metallicity stars~\cite{Asplund:2005yt}.
This level of \li6/\li7 should be considered high as it cannot be
explained by cosmic ray production at such low
metallicities~\cite{Prantzos:2005mh}.  However, in the thermal
environment of stellar atmospheres the absorption lines of \li7 and
\li6 are blended together so that observations become extremely
challenging. Indeed, recent works~\cite{Cayrel:2007te} have claimed
that additional convective motion in stellar atmospheres may lead to
an \li6-unrelated distortion of \li7 absorption lines which could be
confused with positive \lisx-detections. The analysis by Cayrel~{\it
  et~al.}~\cite{Cayrel:2007te} favors the latter explanation and sets
a number of upper limits on the \li6/\li7 in several stars of
interest.

We now move to the main topic of our paper: modification of BBN by new
physics. Instead of concentrating on very specific models of particle
physics and/or gravitation, we review different {\it classes} of
models that modify or abandon the main assumptions of SBBN listed in the
Introduction.

\section{NEW PHYSICS AFFECTING THE TIMING OF MAIN BBN EVENTS}

As shown in Figure~\ref{figure1}, most neutrons are incorporated into
\hef\ at a very specific moment: the end of the D bottleneck. The
amount of helium and deuterium that results has a strong dependence on
the $n/p$ ratio that, in turn, translates into a sensitivity to a much
earlier event in the BBN history: the freeze-out of neutron-to-proton
interconversion. Different physical processes may affect the timing of
both events and thereby change the BBN output.

\subsection{BBN with new degrees of freedom}

\begin{figure}[!t]
  \centerline{\includegraphics[width=0.7\textwidth]{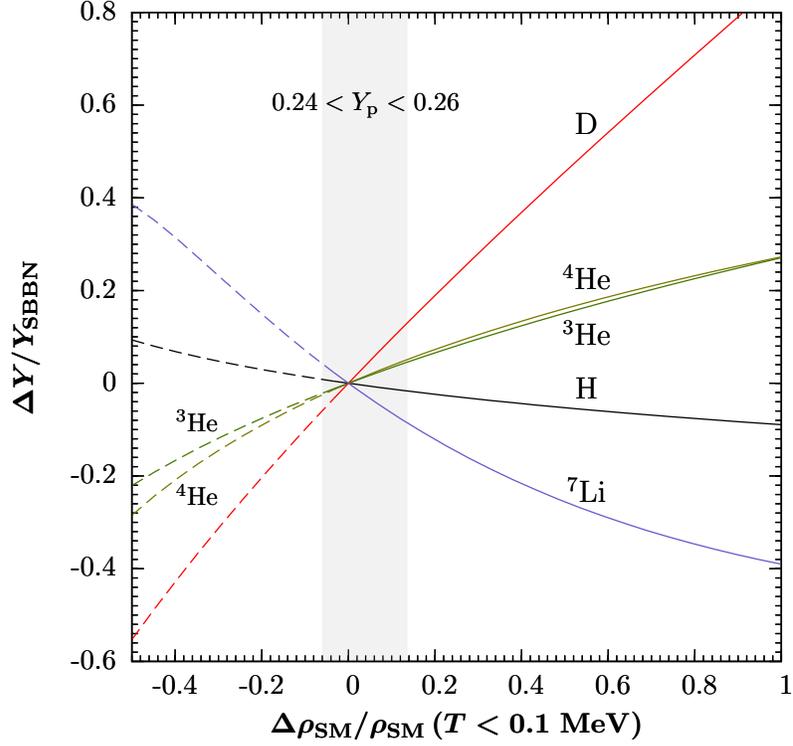}}
  \caption{\base Change of nuclear abundances relative to their SBBN
    values as a function of the ``dark radiation'' component
    $\rho_{dr}/\rho_{\SM}= \Delta\rho_{\SM}/\rho_{\SM}$. The vertical
    band shows the allowed amount of dark radiation that keeps $Y_p$
    in the $0.24\div 0.26$ window. }
\label{figure2}
\end{figure}

A classic application of BBN consists of constraining the number of
the excited relativistic degrees of freedom.  Traditionally, this
procedure is formulated in terms of a constraint on the effective
number of neutrino species,~$N_{\nu,\mathrm{eff}}$.  To be more
general, we re-formulate such a bound as a limit on the amount of
so-called ``dark radiation'',~$\rho_{dr}$. By dark radiation we mean extra
massless or nearly-massless degrees of freedom that have the expected
scaling as the Universe expands, $\rho_{dr} \sim a^{-4}$. Furthermore,
we assume that this additional relativistic component is totally
``passive'' and that it does not exchange energy with any SM species.  This
viewpoint can then be easily applied to the thermally decoupled extra
bosonic or fermionic degrees of freedom.  The addition of dark
radiation to the cosmic energy budget amplifies the Hubble expansion
rate,
\begin{equation} 
  H_{\SBBN} \to H=H_{\SBBN}\sqrt{1+{ \rho_{dr}}/{\rho_{\SM}}},
\end{equation}
leading to an {\it earlier} freeze-out of the $n/p$ ratio and thus to
a larger helium mass fraction. In this formula, $\rho_{\SM}$ and
$H_{\SBBN}$ are the SBBN energy density and the Hubble rate,
respectively. Figure~\ref{figure2} illustrates the changes to the
abundances of helium, deuterium and lithium as a function of
$\rho_{dr}/\rho_{\SM}$ and shows that even the somewhat generous
assumption on the error bar of $Y_p$ translates into a rather tight
constraint on $\rho_{dr}$. One cannot achieve any significant degree
of \lisv\ suppression without violating other bounds.  Figure 2
formally extends to the region of negative $\rho_{dr}$. This may seem
unphysical, but a negative $\rho_{dr}$ effect can be mimicked by
models with decreased neutrino temperatures due to, for instance,
additional electromagnetic energy injection at approximately $T\sim 2$
MeV.  Assuming that \hef\ is limited to the range between $0.24$ and
$0.26$, one arrives at the conclusion that
\begin{equation}
  \label{rhodr}
  -0.06 < \frac{ \rho_{dr}}{\rho_{SM}} < 0.14 .
\end{equation} 
Presented in a traditional way, this constraint corresponds to $
2.6\simle N_{\nu,\mathrm{eff}}\simle 4$.

The constraint on $\rho_{dr}$ can be further specified to models at
hand.  It can be applied to limit the energy density carried by very
light and thermally decoupled scalar fields. For example, if the dark
energy sector at the time of BBN corresponds to a very light classical
scalar field $\phi$ that tracks $\rho_{\SM}$
\cite{Ratra:1987rm,Wetterich:1987fm}, Eq.~(\ref{rhodr}) constrains the
parameters of the potential, $V(\phi)$. For an exponential potential
of the form $V(\phi) = M^4\exp\{-\lambda \phi/M_{Pl}\}$, the energy
density carried by the tracker field is $\rho_{\phi}/\rho_{total} =
4/\lambda^2$, so that the magnitude of $\lambda$ should be larger
approximately five so as to satisfy (\ref{rhodr}).

The result (\ref{rhodr}) can also be developed further into
constraints on the {\it properties} of the dark radiation fields and
the temperature of their thermal decoupling from the SM field content.
Suppose that the dark radiation sector contains $N$ bosonic light
neutral fields that are connected to the SM via some heavy
$\Lambda_n$-scale suppressed operators of dimension $n>4$. Then, the
thermalization rate of dark radiation with the SM particles would
typically scale as $\Gamma_{th} \sim T(T/\Lambda_n)^{2(n-4)}$. The
temperature of thermal decoupling is given by the condition $H(T_d)
\sim \Gamma_{th}$. Therefore, for every given $N$ that threatens to
violate the bound of Eq.~(\ref{rhodr}) one can determine a {\it
  minimal} decoupling temperature, and for any given $n$ place a {\it
  lower} bound on the high-energy scale $\Lambda_{n}$.  For example,
if $N\ge 3$, then the thermal decoupling of these species must occur
around the quantum chromodynamics hadronization epoch ($T\sim
200\,\MeV$) or earlier. By comparing the Hubble rate with the
thermalization rates at that epoch, one arrives at
\begin{equation}
  N\ge 3 ~~\Rightarrow~~ \Lambda_5 > 5\times 10^8~{\rm GeV};~ \Lambda_6 > 5~{\rm TeV}.
\end{equation}
If the interaction is mediated by dimension five operators then the
sensitivity to the $\Lambda$ scale can extend very far---indeed, much
beyond directly accessible energy scales in collider experiments.  In
exactly the same way, one can constrain properties and interactions of
right-handed neutrinos, should they be light.  For example,
Eq. (\ref{rhodr}) can be used to set limits on the strength of the
coupling between left- and right-handed neutrino species. Finally,
even if the right-handed neutrinos are heavy and not excited, BBN
allows one to constrain non-standard properties of the left-handed
neutrino species. For example, a sizable magnetic moment or charge
radius of neutrinos (that correspond to dimension five and six
effective interactions with the external electromagnetic current) may
be strong enough to prolong the thermal coupling of neutrinos to the
electron-positron-photon fluid, thereby leading to essentially higher
$T_\nu$ than predicted by the standard scenario~(\ref{Tnu}). More
discussions on the status of BBN with extra degrees of freedom can be
found in earlier reviews
\cite{Malaney:1993ah,Sarkar:1995dd,Iocco:2008va,Steigman:2007xt}.

\subsection{BBN with sliding couplings and mass scales.}

An alternative, and more exotic way to affect the timing of main BBN
events is realized in models that predict a time evolution of coupling
constants and mass scales.  A tractable version of such models
represents a Brans-Dicke-type scalar field that couples to the SM
operators $F_{\mu\nu}^2$ and  $m_q \bar qq$
where $m_q$ is  a quark mass.
The evolution of the scalar field creates the effect of changing
$m_q$, electric charge, $\Lambda_{QCD}$, Higgs vacuum expectation
value, and so forth.  These changes, in turn, induce changes in the
reaction rates, nuclear binding, and the position of resonances.  Much
effort has been devoted to calculations that make such connections
explicit \cite{Flambaum:2002de,Flambaum:2007mj}. Several works have
addressed the question of coupling variability in connection to BBN;
for the most recent accounts, we refer the reader to, for
instance~\cite{Nollett:2002da,Dmitriev:2003qq,Coc:2006sx,Dent:2007zu}.

\begin{figure}[!t]

\centerline{\includegraphics[width=0.7\textwidth]{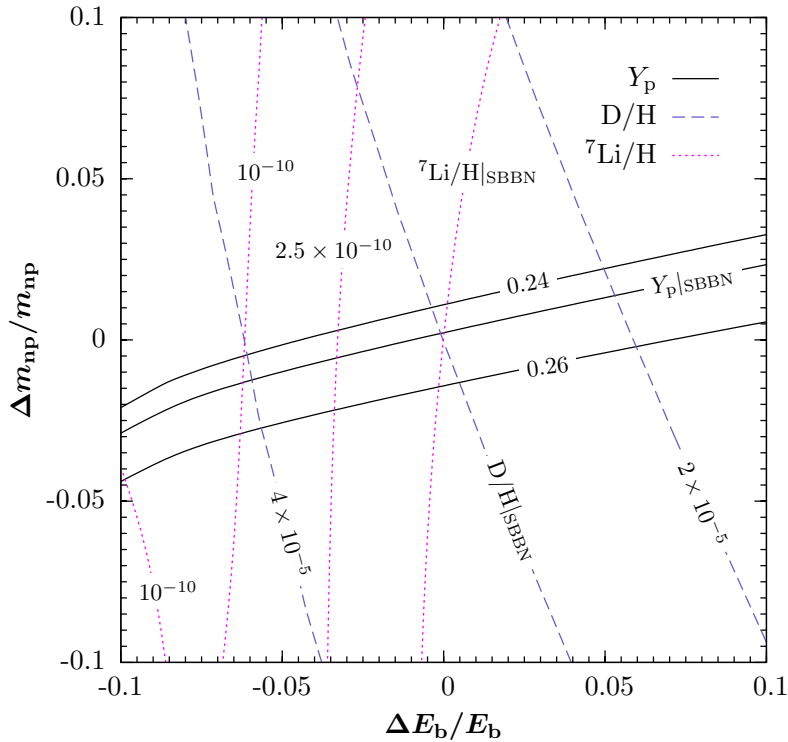}}
\caption{\base Contours of $Y_p$, D/H and \lisv/H are plotted in the
  parameter space of variable neutron-proton mass difference $\Delta
  m_{np}$ and deuteron binding energy $E_d$, normalized to their
  current values.  The ~5-10\% downward change in $E_b$ can
  significantly reduce \lisv\ abundance.  }
\label{figure3}
\end{figure}

Rather than delve into the intricacies of coupling and the scale
dependence of nuclear parameters, we limit ourselves to the following
simplified framework. We assume that the primary effect of the change
of couplings is on the value of the neutron-proton mass difference
$\Delta m_{np}$ and on the deuteron binding energy $E_{d}$.  We
disregard other possible changes, {\it i.e.} in the Coulomb barrier
penetration, shifts in the position of nuclear resonances etc.  Such
an approximation is somewhat justified because both, $\Delta m_{np}$
and $E_d$, enter in the exponents that control (a) the $n/p$ ratio
around its freeze-out time and (b) the D abundance around the end of
the bottleneck.  By using this approach, we are certain to capture the
main effects related to the change of couplings because $\Delta
m_{np}$ influences the $n/p$ freeze-out and the value of $E_d$
dictates the end of the D bottleneck.  Moreover, the deuteron binding
energy is among the most sensitive nuclear parameters to the variation
of $m_q/\Lambda_{QCD}$ due to a strong cancellation of $m_q$-dependent
against $m_q$-independent components in the deuteron binding energy
\cite{Dmitriev:2003qq}.

Figure~\ref{figure3} shows contours of D, \hef\ and \lisv\ abundances
in the $\Delta m_{np}$--$E_d$ plane when both variations are
implemented simultaneously.  The abundances of elements are very
sensitive to the variations of these quantities: helium in particular
is extremely sensitive to $\Delta m_{np}$.  Moreover, among all
elements, lithium appears to be the most sensitive to the variation of
the deuteron binding energy because for smaller values of $E_d$, the
end of the D bottleneck is delayed, and the $\het(\alpha,\gamma)\bes$
reaction is even less efficient than in SBBN.  The possibility of
suppressing the lithium abundance while maintaining the agreement with
helium and deuterium was noted in recent
papers~\cite{Dmitriev:2003qq,Coc:2006sx,Dent:2007zu}. A rather strong
sensitivity of $E_d$ to the $m_q/\Lambda_{QCD}$ parameter ensures that
even a small variation of the latter leads to a drastic change in
lithium.

\section{NON-EQUILIBRIUM BBN}

One of the main assumptions of standard BBN is that all reactions
occur between partners that are perfectly thermalized with the photon
background fluid.  However, even the standard nuclear processes that
occur during BBN lead to a non-thermal energy component which is
released when mass is converted into nuclear binding energy. It is
easy to estimate how much of this energy is injected. Because the
total binding energy of helium is 28 MeV, the total energy release is
just under 2 MeV per nucleon. Although such an energy injection is
noticeable when it occurs  later in the history of the Universe, the
release of such energy at $T_9=0.85$ does not lead to a change in
SBBN predictions (for an analysis of some SBBN non-equilibrium
reactions see Ref. \cite{Voronchev:2008zz}).

More drastically, non-standard decaying or annihilating particles can
also increase the energy release per nucleon. Indeed, if this occurs
predominantly after nucleosynthesis is complete it can lead to strong
departures from the observed pattern of primordial
abundances. Therefore, BBN provides us with a significant sensitivity
to this class of models even when the total energy density stored in
the decaying and annihilating species is completely subdominant to the
energy density of the Universe.  This issue has taken on a great deal
of importance due to its potential connection to particle dark matter,
in which weakly interacting massive particles (WIMPs) can either source
non-equilibrium BBN via their annihilation, or be produced in the
decays of some metastable parent particles.

BBN scenarios with additional energy injection have received a plenty
of attention since their inception
\cite{Ellis:1984er,Levitan:1988au,Dimopoulos:1987fz,Reno:1987qw,
  Dimopoulos:1988ue,Ellis:1990nb,Khlopov:1993ye,Kawasaki:1994sc}, more
recent and more detailed treatments can be found in
\cite{Jedamzik:2004er,Kawasaki:2004yh,Kawasaki:2004qu,Cyburt:2006uv,Cyburt:2009pg}.
By accounting for the qualitative differences in the abundance
signatures, one may distinguish between electromagnetic decays to
$\gamma$s, $e^{\pm}$, and possibly other leptons and decays to hadronic
final states that lead to extra energetic nucleons. Whereas
electromagnetic decays have a significant impact on BBN only at late
times ($\tau\simge 10^5\,$s), after all reactions are effectively
frozen, the hadronic decays may have observable consequences even if
they occur as early as few seconds.

\subsection{BBN with electromagnetic and hadronic energy injection}

When the decaying particle produces mostly electromagnetic radiation,
the treatment of non-equilibrium BBN is relatively simple. Because the
density of the early Universe is quite significant, decaying particles
quickly yield electromagnetic showers. As a result, a potentially very
large energy release per decaying particle ({\it i.e.} $\mathcal{O}(1\
\TeV)$ can be transferred to a large number of
$\Orderof{10\,\MeV}$-energy photons, some of which may have a chance
of interacting with and/or disintegrating light nuclei before their
energy is further dispersed and thermalized.

The main physical process that regulates the \textit{maximal} energy
of particles in the shower is the pair-production in the scattering of
energetic $\gamma$s on thermal photons, {\it i.e.} $\gamma
+\gamma_{T}\to e^- + e^+$. This leads to a so-called
``zeroth-generation'' differential photon spectrum in the form of a
broken power law~\cite{Protheroe:1994dt}:
\begin{align}
  p_{\gamma}(E_{\gamma}) = 
\begin{cases} 
K_0 (E_{\gamma}/E_{\mathrm{low}})^{-1.5}  &\mathrm{for}\
  E_{\gamma}<E_{\mathrm{low}} \\%[-0.3cm]
 K_0 (E_{\gamma}/E_{\mathrm{low}})^{-2.0} & \mathrm{for}\
  E_{\mathrm{low}}<E_{\gamma}<E_C \\%[-0.3cm]
\qquad\quad 0 & \mathrm{for}\ E_{\gamma}>E_C
\end{cases}
\label{eq:pgamma}
\end{align}
where the power break occurs at $E_{\rm low} \simeq m_e^2/(80T)$, and
the spectrum is cut-off at the threshold of pair production $E_C
\simeq m_e^2/22T$~\cite{Kawasaki:1994sc}. The overall normalization
constant $K_0$ of the spectrum is determined by requiring that the
primary injected (electromagnetic) energy $E_0$ be carried by the
photon cascade, that is,~$E_0 = \int d E_\gamma E_\gamma p_\gamma $.

The ansatz~(\ref{eq:pgamma}) for the spectrum immediately tells us the
temperature and time of injection that allow for the
photodisintegration of a certain element. The highest temperature
$T_{\rm ph}$ (one-to-one related to the earliest cosmological time) at
which photodisintegration can occur can be determined by equating
$E_C$ to the nuclear binding energies $E_{b}$ against
photodissociation:
\begin{equation}
  T_{\rm ph} \simeq 
\left\{ \begin{array}{lll}
7\,\keV &  \mathrm{for}\ \bes+\gamma\to\het+\hef  & (E_b = 1.59\,\MeV)\\%[-0.3cm]
5\,\keV &  \mathrm{for}\ \deut+\gamma\to n+ p & (E_b = 2.22\,\MeV)\\%[-0.3cm]
0.6\,\keV & \mathrm{for}\ \hef+\gamma\to \het(\trit)+n(p) & (E_b \simeq 20\,\MeV)
\end{array}\right .
\end{equation}
Once the temperature is so low that the photodisintegration of \hef\
can occur, net {\it production} of D and \het\ becomes possible. An
amount of \li6 can be produced either via photodissociation of \bes\
and \lisv\ or through secondary interactions of the products of \hef\
destruction. This important mechanism is discussed in some detail
later.

\begin{figure}[!t]
\centerline{\includegraphics[width=0.7\textwidth]{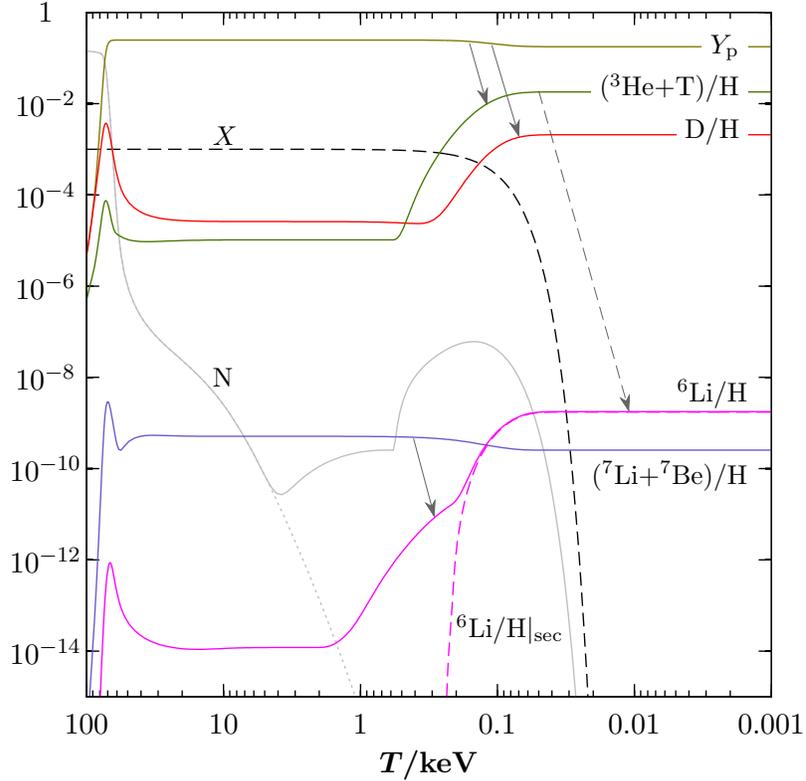}}
\caption{\base Consequences of late decays of a heavy 1~\TeV\ mass
  particle $X$ that releases half of its rest mass in the form of
  electromagnetic energy. The threshold of \hef\ disintegration is
  clearly visible below 1~keV. Primary abundance flows are indicated
  by solid arrows whereas the dashed arrow indicates the secondary
  transformation of $A=3$ nuclei into \lisx. The model is excluded by
  the overproduction of D, \het, and \lisx.  }
\label{figure4}
\end{figure}

The photons in the cascade~(\ref{eq:pgamma}) undergo further
degradation via the (slower) processes of Compton scattering,
pair-production on nuclei, and elastic $\gamma$-$\gamma$ scattering so
that the total number of energetic photons is given by competition
between the injection (decay or annihilation) rate $\Gamma_{\rm inj}$
and the total energy loss rate $\Gamma_{\gamma}(E_\gamma)$.  The
energy spectrum can then be obtained in form of a quasi-static
equilibrium solution~\cite{Cyburt:2002uv}
\begin{equation}
f_{\gamma}^{\rm qse} = n_X\fr{\Gamma_{\rm inj}p_\gamma(E_\gamma)}{\Gamma_\gamma(E_\gamma)},
\end{equation}
where $n_X$ is the time-dependent number density of the decaying and
annihilating particles, $\Gamma_{\rm inj} = \tau_X^{-1}$ for decays,
and $\Gamma_{\rm inj} = \frac12n_X\langle \sigma_{\rm ann}v \rangle $
for self-annihilation.  Depending on the temperature of the primordial
plasma at the time of energy injection, both production and
destruction of elements may occur. One can incorporate these
possibilities into an additional set of Boltzmann equations, which
include the non-thermal photon population and abundances of nuclei
denoted here as $T,A $, and $P$ ($A_T > A_A > A_P$):
\begin{eqnarray}
  -HT\frac{dY_A}{dT} =  \sum_{T} Y_T \int_0^\infty dE_\gamma f_{\gamma}^{\rm qse}(E_\gamma)
  \sigma_{\gamma +T \to A}(E_\gamma) \nonumber \\ 
  - Y_A \sum_{P} \int_0^\infty dE_\gamma f_{\gamma}^{\rm qse}(E_\gamma)
  \sigma_{\gamma +A \to P}(E_\gamma).
\end{eqnarray} 
The solution to this set of equations constrains the amount and timing
of deposited electromagnetic energy.  Figure~\ref{figure4} shows the
results of a sample calculation for the model with a decaying particle
with a lifetime of $\tau_X = 10^8\,$s and an initial energy density
relative to baryons of $m_Xn_X/(m_pn_b)=m_X Y_X/m_p \simeq 1$; there
is an extra assumption that half of the rest mass of the species,
$m_{X} = 1\,\TeV$, is released in the form of electromagnetic energy
into the thermal bath.  One can see the significant increase in \het,
D, and \lisx\ abundances. The solid arrows indicate the main
transformations of the elemental abundances under the influence of the
dissociating radiation.  The model considered in this graph is in
stark conflict with observations and is therefore excluded.

It is intriguing to investigate whether the injection of
electromagnetic energy may reduce the abundance of \lisv+\bes. A
dedicated study~\cite{Ellis:2005ii} found that such a reduction is
generally difficult to achieve because either D is also destroyed or
\het/D is overproduced beyond observationally acceptable levels.
Although the overall conclusion is negative, \textit{i.e.} the
electromagnetic destruction of \lisv\ is problematic, we stress that
``just-so'' solutions can always be found.  For example, consider a
relatively light but abundant neutral particle with mass $m$ in the
narrow range between $\sim 3.5$ and $4.5$ MeV. \bes\ could indeed be
destroyed while D, \het, and \lisx\ remain unaffected because the
maximum released energy, $m/2$, lies below their photo-disintegration
threshold.

The treatment of energy injection via hadronic channels is
significantly more complicated because additional effects must be
taken into account.  Partons emitted in a decay or annihilation are
quickly hadronized and the highly energetic fragmentation products
[mostly pions but also (anti-)nucleons] are released into the
plasma. Only long-lived mesons---namely, charged pions~$(\pi^\pm)$ and
kaons ($K^\pm,\, K^0_{\mathrm{L}}$)---with lifetimes of
$\Orderof{10^{-8}\, {\rm s}}$ have a chance to interact with
background nuclei before decaying.
A very early hadronic energy injection ($300\,\keV\simle T\simle
1\,\MeV$) allows the $\pi$-mesons to participate in charge exchange
reactions, $\pi^{-}+p\to\pi^0+n$. Such reactions lead to a higher
$n/p$ freeze-out value, thereby increasing the helium mass fraction
$Y_p$.  Similar effects can be caused by anti-nucleons, which tend to
annihilate predominantly on protons. The most important difference
with respect to electromagnetic energy releases comes from the
injection of energetic nucleons after the formation of \hef.  The
propagation of energetic nucleons through the primordial plasma causes
spallation processes on \hef. Reactions caused by energetic
nucleons---$n +\hef\to \trit + p + n $, $n +\hef\to \deut + p + 2n $
and others---in turn generate energetic $A=3$ elements that can
participate in endo-thermic nuclear reactions that are forbidden in
SBBN.

What determines the efficiency of spallation processes and of
subsequent secondary non-thermal reactions is the rate at which the
energetic charged nuclei and neutrons are stopped by the primordial
plasma. The dominant thermalization processes for charged nuclei are
Coulomb interactions with $e^{\pm}$ and Thomson scattering off thermal
photons. Below the $e^+e^-$ annihilation threshold the stopping power
of the plasma rapidly drops with temperature, reaching its minimum at
approximately $T\simeq 20$ keV. The energy loss of the neutrons
initially occurs via its magnetic-moment interaction with electrons
and positrons. At later stages ($T\simle 80\,$keV), neutrons loose
their energy dominantly by scattering on protons and \hef.

The resulting constraints on energy injection are described in great
detail elsewhere
\cite{Jedamzik:2004er,Kawasaki:2004yh,Kawasaki:2004qu,Cyburt:2006uv,Cyburt:2009pg}.
BBN results are most sensitive to the energy injection after $10^8$ s,
where constraints as strong as $\sim 1$ MeV per nucleon result from
the secondary \lisx\ production, and do not generally depend on the
branching ratio to hadrons.  Constraints on electromagnetic energy
injection depend, to a large extend, on whether \hef\ can be
photo-dissociated, and they become much weaker if energy is released
when the temperature is above 0.6 keV.  Hadronic decays allow one to
probe much earlier times. For fully hadronic decays to energetic
$q\bar q$ pairs, similar sensitivity (1 MeV per nucleon) applies to
injection at much earlier times, specifically $t\sim 10^4$ s.  Perhaps
one of the most interesting features of energy injection with nucleons
in the final state is the effect of extra neutrons at $T\sim
40\,\keV$, which may lead to an important depletion of the total
lithium abundance.

\subsubsection{Neutron excess at 40 keV and suppression of \lisv.}

\begin{figure}[!t]
\centerline{\includegraphics[width=0.7\textwidth]{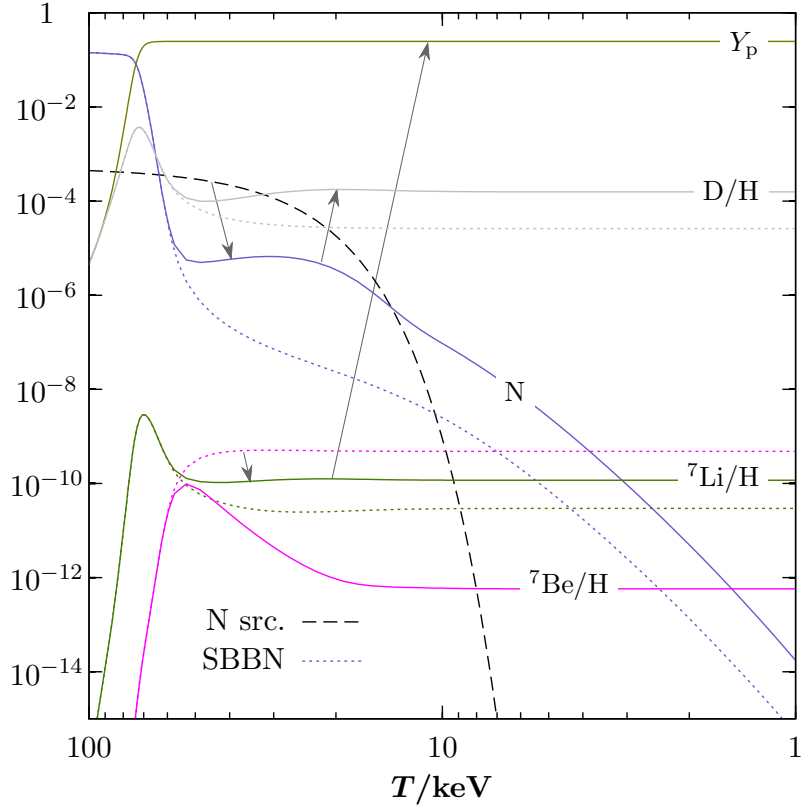}}
\caption{\base Effects on element abundances in response to an
  elevated (thermal) neutron content, sourced by a decaying species,
  at early times. Although neutrons are mostly incorporated into D,
  they also transfer \bes\ nuclei into the \lisv\ reservoir from which
  they are more susceptible to proton burning. This mechanism
  suppresses the overall outcome of the $A=7$ elements.}
\label{figure5}
\end{figure}

An interesting aspect of non-thermal BBN with hadronic decays and
annihilations is the possibility of alleviating the tension between
the Spite plateau value and the predicted abundance of \lisv, {\it
  e.g.}  ``solve the \lisv\ problem''. In SBBN the \bes\ abundance is
controlled by the $\het+\hef\to \bes+\gamma$ reaction and by the
combination of the two reactions that destroy it:
\begin{eqnarray}
n+{^7\rm Be}\to p +{^7\rm Li};~~~
\,\, p+{^7\rm Li}\to {^4\rm He} + {^4\rm He}\, .
\label{Li}
\end{eqnarray}
The increase of neutrons in a narrow temperature interval $60\,
\keV\simge T\simge 30\, \keV$ (in other words, during or just after
\bes\ synthesis) may amplify the efficiency of this destruction.  The
injection of $\simge 10^{-5}$ neutrons per baryon, regardless the
microscopic cause of such an injection, may enhance \bes$\to$\lisv\
and lead to an overall depletion of \bes+\lisv\
\cite{Reno:1987qw,Jedamzik:2004er}. When the temperature drops below
30 keV, the \lisv\ burning reaction drops out of equilibrium, the
lithium destruction stops, and the supply of ``extra neutrons'' has no
further effects on the overall \bes+\lisv\ abundance.  At the same
time, most of the extra neutrons injected around the time of BBN are
removed by the radiative recombination with protons leading to the
generation of extra D.  Therefore, one should expect that this
mechanism of depleting \bes\ is tightly constrained by the abundance
of D. The temperature evolution of elemental abundances in the
presence of an extra source of thermal neutrons is illustrated in
Fig.~\ref{figure5}.

\begin{figure}[!t]
\centerline{\includegraphics[width=0.7\textwidth]{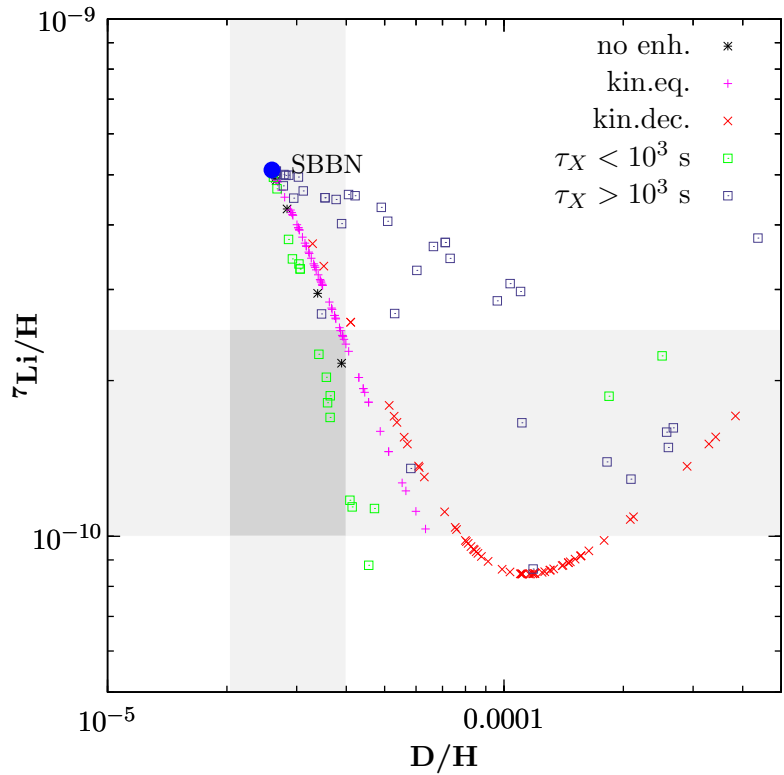}}
\caption{\base \lisv\ versus D abundance in the BBN scenario with
  ``extra'' neutrons. The overall number of injected neutrons, as well
  as the temporal patterns of injection, is varied. Patterns
  consistent with particle decays and annihilation, both with and
  without Sommerfeld enhancement, are considered.  }
\label{figure6}
\end{figure}

Elevated neutron concentrations can be caused by decaying or
annihilating particles that contain a significant number of baryons in
the final state. This presence is amplified when the energetic
particles lose their energy in collisions with the primordial
plasma. Many details of such processes depend on concrete particle
physics realizations: particle masses, abundances, lifetimes,
branching ratios to nucleons, and so forth all play important
roles. In Figure \ref{figure6} we both emulate the process of energy
injection by assuming very rapid thermalization and study the effect
of extra neutrons on \lisv+\bes\ and D abundances by varying the
amount of injected neutrons, while assuming different time (or
temperature) patterns for such injections.  Specifically, we include
cases of decaying particles (with different lifetimes), and
annihilating WIMPs, with or without possible Sommerfeld or resonant
enhancement of annihilation \cite{Pospelov:2008jd}.  If not carefully
tuned, the injection of neutrons that reduces the total \lisv\
abundance by a factor of two, would over-predict the D/H ratio. Only a
relatively small subset of models, typically those whose decaying
particles have lifetimes below $1000$ s are consistent with D/H, and
provide enough suppression for \lisv. However, the injection of
energetic neutrons, ignored in this treatment is also important
because of secondary and tertiary reactions induced by the
neutrons. These reactions lead to an increase of rare light elements
such as \lisx\ and \ben. This method of correcting the lithium problem
is subject to additional constraints from \lisx\ and \ben.

\subsubsection{Production of \li6, \ben, and $^{10}$B in secondary  collisions.}

As stated in the introduction, the abundances of these elements
predicted by SBBN are extremely low. The path to them is heavily
hindered as it is guarded by (a) the breaks at $A=5$ and 8, (b) the
low efficiency of the SBBN reaction (\ref{He4DLi6}) and (c) the
efficient destruction of these elements by protons which stops only at
$T\sim 10$ keV when the destruction rates fall below the Hubble
expansion rate.  Non-equilibrium BBN provides an important mechanism
for circumventing at least some of that suppression through access to
endo-thermic reactions that have much higher cross sections
\cite{Dimopoulos:1987fz}.

\begin{figure}[!t]
\centerline{\includegraphics[width=0.7\textwidth]{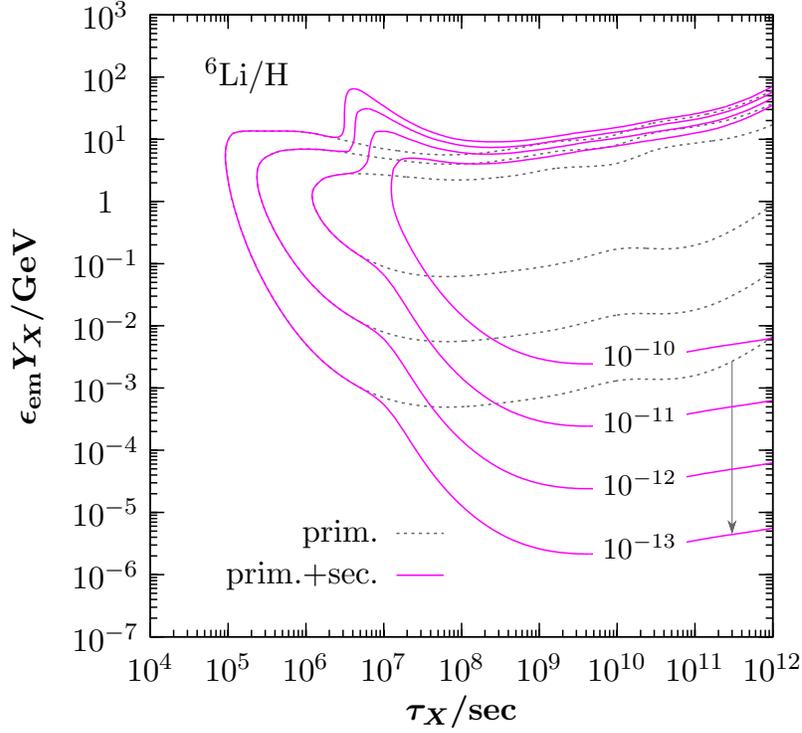}}
\caption{\base Contour lines of \lisx/H in the plane of X-lifetime
  $\tau_X$ and total injected (electromagnetic) energy per baryon:
  $\epsilon_{\mathrm{em}}Y_X$. Only below $\sim 10\,\keV$ ($t>
  10^4\,$s) are \lisv\ and \bes\ dissociated by energetic photons
  producing ``primary'' \lisx\ (dotted lines). Additionally, \lisx\ is
  fused via the processes of Eq.~\ref{nonthermal}).  The importance of
  this ``secondary'' production at late times represented by the
  vertical arrow which indicates the significant elevation of total
  \lisx/H (solid lines) with respect to the primary yield (dotted
  lines).}
\label{figure7}
\end{figure}

Energetic mass three nuclei, \het\ and tritium, produced via
electromagnetic or hadronic energy injection ({\it i.e.} via
spallation or photodisintegration) collide with $\alpha$ particles
from the thermal bath. Such secondary collisions produce \lisx\ via
the following set of reactions.
\begin{eqnarray}
^3{\rm H} + \hef \to \lisx + n, ~~  Q = -4.78{\rm MeV};
\nonumber
\\\label{nonthermal}\!\!\!\!\!\!\!\!\!\!\!\!\!
\het + \hef \to\lisx +p, ~~ Q = -4.02{\rm MeV};
\\
\nonumber
^3{\rm H} + \hef \to \hesm + p, ~~  Q = -7.5{\rm MeV},
\end{eqnarray}
where $^6{\rm He}$ subsequently $\beta$-decays to \lisx.  For energies
of projectiles that are approximately 10 MeV, the cross sections for
these non-thermal processes are on the order of 100 mb---$10^{7}$
times larger than the SBBN cross section for the production of
\lisx. The overall efficiency of producing \lisx\ from energetic \het\
and tritium reaches its maximum at $\Orderof{10^{-4}-10^{-3}}$ below
$T=10$ keV; therefore, even one energetic nucleus of tritium per one
million protons can have a very significant impact on \lisx. This
enhancement figure underscores the sensitivity of \lisx\ to
non-thermal BBN and makes it an important probe of energy injection
mechanisms in the early Universe.

Figure~\ref{figure7} illustrates the importance of accounting for
secondary reactions of the kind (\ref{nonthermal}) when tracking the
\lisx\ output in scenarios with electromagnetic energy release at late
times. Shown are contour lines of constant \lisx/H for varying
lifetime $\tau_X$ of an decaying species $X$ and different total
injected (electromagnetic) energy per baryon,
$\epsilon_{\mathrm{em}}Y_X$. Whereas the dotted lines represent the
``primary'' abundance of \lisx\ generated in the photodissociation of
\lisv\ and \bes, its ``secon\-dary'' production becomes very important
as soon as $\tau_X\simge 10^7\,$s. This is highlighted by the
vertical arrow indicating the increased sensitivity to a decaying
species by three orders of magnitude, attributed to the secondary
production mechanism. In our treatment of photodissociation processes
we primarily follow the approach of Ref.~\cite{Cyburt:2002uv}.

The doubtful observational status of \lisx\ and the possibility of its
stellar depletion over and above the depletion of \lisv\ should
stimulate studies of \ben\ and boron production via the non-thermal
nucleosynthesis. So far, this subject has escaped the attention of the
groups working on non-equilibrium BBN. Note that {\it tertiary
  processes} may be responsible for the non-equilibrium \ben\
production.  The $^6{\rm He}$ and \lisx\ nuclei emerging from
secondary reactions (\ref{nonthermal}) can collide further with
particles from the bath,
\begin{eqnarray}
  \lisx +\hef \to \benm + p, ~~ Q = -2.13 {\rm MeV}
  \\
  \nonumber
  ^6{\rm He} +\hef \to \benm + n, ~~ Q = 0.60{\rm MeV} ,
\end{eqnarray}
and generate \ben\ from $A=6$ nuclei with similar efficiencies to the
$A=3\to A=6$ transition. The actual output of \ben\ relative to \lisx\
is enhanced beyond a crude estimate of $\Orderof{10^{-4}-10^{-3}}$ if,
for example, the energy injection occurs both at very early times and
above $T=12$ keV, where \lisx\ is more rapidly destroyed by proton
reactions than \ben\ is.  Given the absence of a primordial \ben\
plateau down to the level of $10^{-13}$, it is desirable to
investigate the tertiary production mechanisms of beryllium and boron
in more detail~\cite{PospelovPradler}.

\subsection{Energy injection by WIMPs}

Decays of heavy relics may represent the simplest possible way to
achieve energy injection.  However, without specifying the model and
the physical mechanisms that lead to the production of unstable
particles and their subsequent decay, one cannot estimate the amount
or the timing of the energy injection.  In contrast, WIMPs represent a
somewhat more restricted framework in which the leading mechanisms
regulating the WIMP abundance at BBN time is their self-annihilation
at earlier times. Assuming a weak-scale mass for a relic WIMP as well
as a typical range for the annihilation cross section, one can deduce
its abundance as a function of mass and annihilation rate.  The most
restrictive framework is that of stable neutral WIMPs that form the
dominant component of cold dark matter. Although most of the WIMPs
annihilate when the temperature drops below the WIMP mass, a residual
annihilation persists even at BBN times~\cite{Jedamzik:2004ip}. The
question of WIMP annihilation at early times, during BBN, is important
in view of the attempts to detect signatures of WIMP annihilation in
our own Galaxy. Indeed, a broad similarity among WIMP velocities
inside the Galaxy and during BBN tells us that the kinematics of WIMP
collisions must be very similar.

The most straightforward case to consider is a single species of WIMP
dark matter $X$ with annihilation rate $(\sigma v)_0 \simeq 3\times
10^{-26}{\rm cm^3s^{-1}}$ averaged over the thermal bath at the
freeze-out. Such an annihilation rate yields a total energy density of
WIMPs that is close to what is required for cold dark matter by
observations. This result provides us with a convenient normalization
point for the annihilation rate at \emph{arbitrary} WIMP velocities in
the center-of-mass frame:
\begin{equation}
  \sigma(v) v = S(v)(\sigma v)_0 .
\end{equation}
In the simplest case, in which the annihilation is mediated by
short-distance forces and proceeds in the $s$-wave, the velocity
dependence is trivial: $S(v)=1$.  The fraction of WIMPs that is still
annihilating at the time of BBN during one Hubble time $H^{-1}$ is
given by
\begin{equation}
  f_X(t_{\BBN}) \simeq \left. \frac{\langle\sigma(v)v \rangle n_X }{H}\right|_{t=t_{\BBN}}
\label{eq:fX}    
\end{equation}
where $\langle ... \rangle$ denotes a thermal average that can be
calculated once the velocity dependence of $S(v)$ and the thermal
distribution of WIMPs are specified.

We are interested in finding the fraction of annihilating WIMP
particles within a Hubble time at $ T < 10\,\keV$, specifically, at a
temperature scale below which \lisx\ is no longer susceptible to
nuclear burning in the $\lisx(p,\alpha)\het$ reaction.  In general,
this fraction can be expressed via the freeze-out temperature $T_f
\sim 0.05 m_X$ and the number of effective degrees of freedom at an
arbitrary temperature $T<T_f$ \cite{Jedamzik:2009uy}:
\begin{equation}
  f_X(T) \simeq \left[\frac{g(T)}{g(T_{f})}\right]^{1/2}
  \frac{T}{T_{f}}
  \langle S(v) \rangle_T .   
  \label{fX(T)}      
\end{equation}
There are several generic options for the temperature scaling of the
$S$-factor in (\ref{fX(T)}). In the simplest case of $S(v)=1$, one can
use Eq.~(\ref{fX(T)}) to deduce that for a WIMP of mass $M_X =
100\,$GeV ($T_F\sim 5$ GeV) only a small fraction, $f_X\approx 6\times
10^{-7}$, of the $X$-particles has the chance to annihilate at $T
\simeq 10\,$keV and below.  This corresponds to an injection of
approximately $10^{-15}$ GeV per photon.  Nevertheless, even this tiny
fraction may be sufficient to produce a \li6 abundance up to the
$10^{-12}$-level, provided that hadronic annihilation channels
dominate. Lighter WIMPs may have a more pronounced effect on
BBN~\cite{Jedamzik:2004er}. Short-range mediated annihilation in the
$p$-wave gives $S(v) \sim v^2$ and negligible energy release at BBN.

Much enhanced annihilation rates occur in models where interactions
are mediated by an attractive Coulomb-like force and/or have
near-threshold resonances. Both mechanisms of enhancing the galactic
annihilation have been widely discussed (see {\it e.g.}
\cite{Pospelov:2008jd}) in an attempt to link some cosmic-ray
anomalies---such as an elevated positron fraction $e^+/(e^-+e^+)$
observed by the PAMELA satellite experiment~\cite{Adriani:2008zr}---to
dark matter annihilation. For example, an attractive $-\alpha'/r$
potential in the WIMP sector leads to a significant enhancement of
annihilation at low velocities via a well-known Sommerfeld factor
$\sigma v \sim (\pi \alpha'/v)[1-\exp(-\pi \alpha'/v)]^{-1}$; when
$v\simle \pi\alpha'$, $S(v)\simeq \pi\alpha'/v$. This enhancement, in
turn, generates a $\sim T^{-1/2}$ scaling of $\langle S(v)
\rangle_{T}$ in (\ref{fX(T)}) when WIMPs are still in kinetic
equilibrium with the plasma. Once the WIMPs' interactions with the
thermalized plasma species have ceased, $\langle S(v) \rangle_{T}$
falls even more rapidly as $\sim T^{-1}$. Therefore, the fraction of
WIMPs annihilated in a Hubble time scales with photon temperature as
follows:
\begin{align*}
  \text{\parbox{3cm}{Sommerfeld\\[-0.1cm]enhancement:}}
  \begin{cases}
   f_X(T) \sim T^{1/2} & \quad\text{kinetic equilibrium}\\
   f_X(T) \sim {\rm const} & \quad\text{kinetically decoupled}
  \end{cases}
\end{align*}
Similarly, the presence of narrow resonances just above the $XX$
annihilation threshold may drastically boost the annihilation at low
energies. For a narrow resonance at some energy $E_R$ above the
di-WIMP threshold the pattern of energy injection has a sharp cutoff:
\begin{align*}
  \text{\parbox{3cm}{Resonant\\[-0.1cm]enhancement:}}
  \begin{cases}
   f_X(T) \sim \exp(-E_R/T) & \quad\text{kinetic equilibrium}\\
   f_X(T) \sim \exp(-E_RT_{kd}/T^2)& \quad\text{kinetically decoupled}
  \end{cases}
\end{align*}
where $T_{kd}$ is the temperature of WIMP kinetic decoupling. Notably,
the last pattern (resonant annihilation, kinetically decoupled WIMPs)
has the same time dependence as the energy injection as particle
decays, $\exp(-t/\tau_X)$, on account of the $t\sim 1/T^2$ relation.
If the annihilation cross sections $\langle \sigma v\rangle({t_\BBN})$
are enhanced by some large factor, $\sim \Orderof{10^{3}}$, relative
to the fiducial WIMP rate $(\sigma v)_0$, then WIMP physics can have a
significant impact on the primordial abundance of \lisx. However, a
large class of models that predict the WIMP-enhancement of the PAMELA
positron signal produce light particles (photons, electrons, light
mesons) via annihilation, which has a much weaker effect on \lisx.

A case of independent interest is the decay of a WIMP-like state with
a stable non-SM particle in the final state. The initial abundance of
so-called ``parent'' WIMPs is given by its annihilation at the
freeze-out, whereas their decays may source the dark matter energy
density.  If the amount of energy released in such decays is on the
order of the parent WIMP mass, then in such a scenario an amount of
energy comparable to the dark matter energy density could be released
into the primordial plasma. If so, BBN with energy injection sets a
rather strict constraint: The lifetime of the parent particle must be
less than $\sim 10^7$~s if the decays are fully electromagnetic and
shorter than $10^4$~s if there is a significant hadronic component
among the decays products. In the next section we explain that the
sensitivity to the lifetime of such parent WIMP particles can be even
stronger if they carry negative electric charge.

\section{CATALYZED BBN}

In this section we discuss changes to BBN in a scenario where
particles from a new physics sector {\it participate} in thermal
nuclear reactions before decaying. Such an unusual situation may arise
if the particles are charged under the electromagnetic or strong group
of the SM~\cite{Pospelov:2006sc}.  The idea of particle physics
catalysis of nuclear reactions dates from the 1950s, when
muon-catalyzed fusion became a subject of active theoretical and
experimental research in nuclear physics.  More recently, interest in
the possibility of nuclear catalysis by hypothetical negatively
charged particles that live long enough to participate in nuclear
reactions at the time of BBN has intensified
\cite{Pospelov:2006sc,Kohri:2006cn,Kaplinghat:2006qr,Cyburt:2006uv,
  Hamaguchi:2007mp,Bird:2007ge,Jittoh:2007fr,Jedamzik:2007cp,Jedamzik:2007qk,
  Pospelov:2007js,Kusakabe:2007fv,Pospelov:2008ta,Kamimura:2008fx}
(see also
Ref.~\cite{DeRujula:1989fe,Dimopoulos:1989hk,Rafelski:1989pz} for
earlier work on the subject).  The essence of the idea is very simple:
A negatively charged massive particle, which we term \xm, gets into a
bound state with the nucleus of mass $m_N$ and charge $Z$, forming a
large compound nucleus with charge $Z-1$, mass $M_X+m_N$, and binding
energy in the $\Orderof{0.1-1}$ MeV range.  Once the bound state is
formed, the Coulomb barrier is reduced, signaling higher
``reactivity'' between the compound nucleus and other nuclei. Even
more importantly, new reaction channels may open up and avoid
SBBN-suppressed production mechanisms \cite{Pospelov:2006sc}, {\it
  e.g.} Eq. (\ref{He4DLi6}), thus clearing the path to the synthesis
of very rare isotopes such as \lisx\ and \ben.

Interest in catalyzed BBN (CBBN) is partly motivated by a possible
connection to dark matter.  Even though dark matter should not be
charged, it is possible that it comes with a relatively long-lived
charged counterpart. An example of this kind is weak-scale
supersymmetry (SUSY), where the lightest supersymmetric particle (LSP)
being the gravitino and the next-to-LSP (NLSP) a charged slepton. In
that case the decay of the NLSP is greatly delayed by the smallness of
the gravitino-lepton-slepton coupling $\sim M_{\rm Pl}^{-1}$. Another
example in the same vein is a nearly degenerate stau-neutralino
system, in which case the longevity of the charged stau against the
decay to the dark matter neutralino is ensured as long as the mass
splitting of the stau-neutralino system is below 100 MeV
\cite{Jittoh:2005pq}.  In these two examples, both, the gravitino and
neutralino represent viable dark matter candidates.  A very important
aspect of CBBN is that the abundance of charged particles before they
begin to decay is given by their annihilation rate at freeze-out. In
most models these particles' abundance is easy to calculate. If no
special mechanisms are introduced to boost the annihilation rate, then
the abundance of charged particles per nucleon is not small [ typical
range of $Y_X = n_X/n_b \sim (0.001-0.1) m_X/{\rm TeV}$]. Moreover,
the charged states accompanying dark matter are often required in
order to facilitate the WIMP annihilation process which results in a
cosmologically acceptable dark matter abundance---a phenomenon known
as co-annihilation.

\subsection{Catalysis by stable charged particles.}

In this section we describe BBN catalysis by use of negatively charged
heavy relics $X^-$; we do not attempt to place this phenomenon in a
specific particle physics model framework, and we {\it ignore} the
effects of energy injection due to the $X^-$ decay.  We describe an
elementary particle, so typical collider bounds would require
$m_{X^-}\simge 100$~GeV.

\subsubsection{Properties of the bound states.}

For light nuclei that participate in BBN, we can assume that the
reduced mass of the nucleus-\xm\ system is well approximated by the
nuclear mass, so that the binding energy is given by $Z^2\alpha^2 m_A
/2$ when the Bohr orbit is larger than the nuclear radius. However,
this is a poor approximation for all nuclei heavier than $A=4$, and
the effect of the finite nuclear charge radius must be taken into
account. In Table \ref{table1} we give the binding energies and the
recombination temperature for each bound state. The recombination
temperature is the temperature at which the photodissociation rate of
bound states becomes smaller than the Hubble expansion rate.  Below
these temperatures, bound states are practically stable. The most
important benchmark temperatures for CBBN are $T \sim 30$, 8, and 0.5
keV for the respective bound states (\bes\xm), (\hef\xm), and ($p$\xm)
when they can be formed without efficient suppression by the
photodissociation processes.  Importantly, these properties of the
bound states are generic to any CBBN realization when \xm\ does not
participate in strong interactions; in other words, they are
determined entirely by the charge of \xm\ and electromagnetic
properties of the nuclei and therefore are applicable to SUSY and
non-SUSY models alike. Also, the (\beet\xm) compound nucleus is
stable, which may open the path to synthesis of $A>8$ elements in
CBBN.

\begin{table}[!t]
\begin{center}
\begin{tabular}{cccccccc}
\toprule
bound state &            $a_0$ [fm] &     $|E_b|$ [keV] &~$T_0$ [keV]~ \\ 
\midrule
(p$X^-$)&                 29  &                 25    & 0.6  \\ 
(\hef$X^-$)&            3.63 &        346     &  8.2 \\ 
%\het$X^-$&          299  &   4.81 &    1.76    &       276         &    2.50   &         267     &  6.3 \\ \hline
%\lisx$X^-$&        1343  &   1.61 &    2.22    &       930         &    3.29   &         780     &  19  \\ \hline
%\lisv$X^-$&        1566  &   1.38 &    2.33    &       990         &    3.09   &         870     &  21  \\ \hline
(\bes$X^-$)&            1.03 &          1350    &  32  \\ 
(\beet$X^-$)&           0.91 &           1430    &  34  \\ 
%\hef $X^{--} $ &   1589  &   1.81 &    1.94    &       1200        &      2.16 &         1150    &  28  \\ \hline\hline
\bottomrule
\end{tabular}
\end{center}
\caption{\base Properties of the bound states: Bohr radius $a_0 = 1/(Z\alpha m_N)$, 
  binding energies $E_b$ calculated for realistic charge radii, and 
  ``photo-dissociation decoupling'' temperatures
  $ T_0$.
}
%\vspace{-0.5cm}
\label{table1}
\end{table}

\subsubsection{ Catalysis at 30 keV: suppression of \bes.}

While the Universe cools to temperatures of 30 keV, the abundances of
D, \het, \hef, \bes\ and \lisv\ are already close to their freeze-out
values, although several nuclear processes remain faster than the
Hubble rate. At such temperatures, a negatively charged relic can get
into bound states with \bes\ and form a (\bes\xm) composite object
through the photo-recombination process. Once a composite object is
formed, new destruction mechanisms for \bes\ appear. For models with
weak currents connecting nearly mass-degenerate \xm\ and \xz\ states,
a very fast internal conversion followed by the $p$-destruction of
\lisv\ becomes possible:
\begin{equation}
\bexm\to \lisv\,+X^0;~\lisv+p\to 2\alpha.
\label{intconversion}
\end{equation}
Alternatively, \lisv\ can be destroyed via recombination with \xm\ and
subsequent conversion to unstable $^7$He \cite{Jittoh:2007fr}.  When a
\xm\ $\to$\xz\, weak transition is not allowed, the destruction of
\bes\ can be achieved via the following chain:
\begin{equation}
\label{pburning}
\bexm + p \to (^{8}{\rm B}X^-) +\gamma:~ (^{8}{\rm B}X^-) \to (^{8}{\rm Be}X^-) + e^+\nu,
\end{equation}
which is greatly enhanced by atomic resonances in the $\bexm $ system
\cite{Bird:2007ge}.

The rates for both mechanisms may be faster than the Hubble rate and
may therefore leading to a sizable suppression of \bes\ abundance {\it
  if} (\bes\xm) bound states are forming efficiently.  In other words,
(\bes\xm) serves as a bottleneck for the CBBN depletion of \bes.  The
recombination rate per \bes\ nucleus leading to (\bes\xm) is given by
the product of the recombination cross section and the concentration
of \xm particles.  It is easy to show that for $Y_X \simle 0.01$ the
recombination rate is too slow to lead to a significant depletion of
\bes. Detailed calculations of the recombination rate and numerical
analyses of CBBN at 30 keV \cite{Bird:2007ge,Kusakabe:2007fv} find
that a suppression of \bes\ by a factor of two is possible for (a)
$Y_X \ge 0.1$ if the mechanism in Eq.~(\ref{pburning}) alone is
operative, and (b) for $Y_X \ge 0.02$, if the internal conversion of
Eq.~(\ref{intconversion}) is allowed.  Even if $Y_X \sim
\Orderof{0.1-1}$, \xz\ could still be compatible with dark matter as
long as $m_{X^0}$ is suppressed relative to $m_{X^-}$.

\subsubsection{ Catalysis at 8 keV: enhancement of \lisx\ and \ben.}

As the Universe continues to cool below 10 keV, the efficient
formation of (\hef\xm) bound states becomes possible.  If we make the
reasonable assumption that $Y_{X} \simle Y_{\hef}$ then the rate of
formation of bound states per \xm\ particle is given by the
recombination cross section and the concentration of the helium
nuclei.  A numerical analysis of recombination reveals that at $T
\simeq 5 $~keV approximately 50\% of the available \xm\ particles are
in bound states with \hef\ \cite{Pospelov:2006sc}.

As soon as (\hef\xm) is formed, new reaction channels open up. In particular, 
a photonless thermal production of \lisx\ becomes possible
\begin{equation}
(\hef X^-) + {\rm D} \to \lisx + X^-;~~ Q \simeq 1.13{\rm MeV} .
\label{cbbnLi6}
\end{equation}
This rate exceeds the SBBN production rate by approximately six orders
of magnitude. The production of \ben\ may also be catalyzed, possibly
by many orders of magnitude relative to the SBBN values, with the
following thermal nuclear chain \cite{Pospelov:2007js}:
\begin{eqnarray}
(\hef X^-) + \hef \to (\beetm X^-) + \gamma ;~~ 
(\beetm X^-) + n \to \benm  + X^-.
\label{cbbnBe9}
\end{eqnarray}
Both reactions at these energies are dominated by resonant
contributions, although the efficiency of the second process in
(\ref{cbbnBe9}) is not fully understood.  (A recent
calculation~\cite{Kamimura:2008fx} places the resonance found in the
$(\benm X^-) $ system~\cite{Pospelov:2007js} below the
threshold. However, because this is a reaction with a neutral
particle, even the sub-threshold resonance may give a strong
enhancement of the neutron capture.)

\begin{figure}[!t]
\centerline{\includegraphics[width=0.8\textwidth]{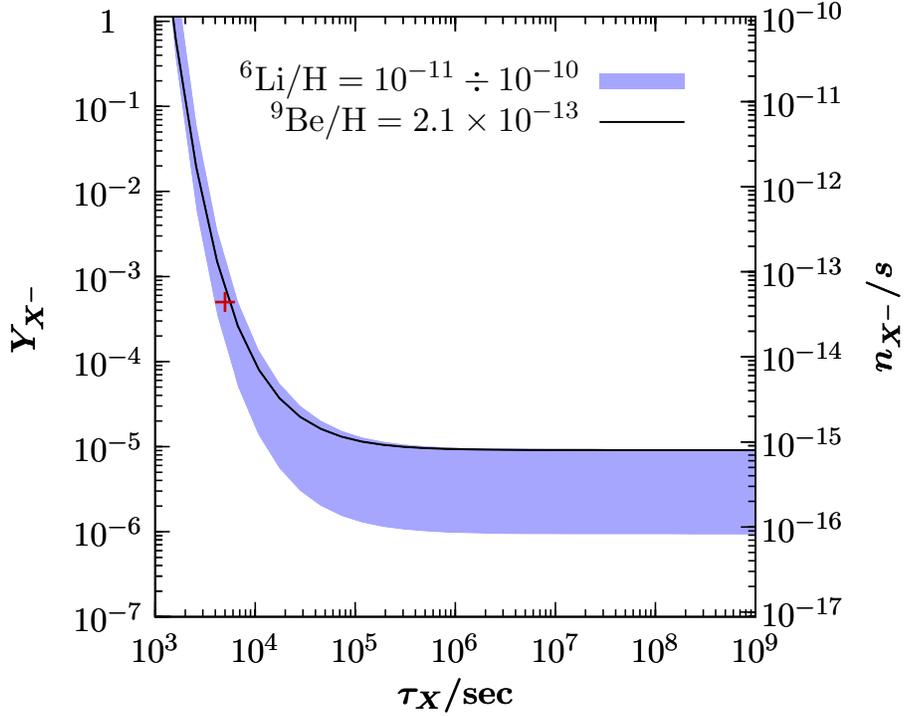}}
\caption{\base CBBN constraints on the abundance versus lifetime of
  \xm.  The red cross corresponds to a point in the parameter space;
  the temporal development of \lisx\ and \ben\ for this point is shown
  in Figure~9~\cite{Pospelov:2008ta}.  }
\label{figure8}
\end{figure}

\begin{figure}[!t]
\centerline{\includegraphics[width=0.7\textwidth]{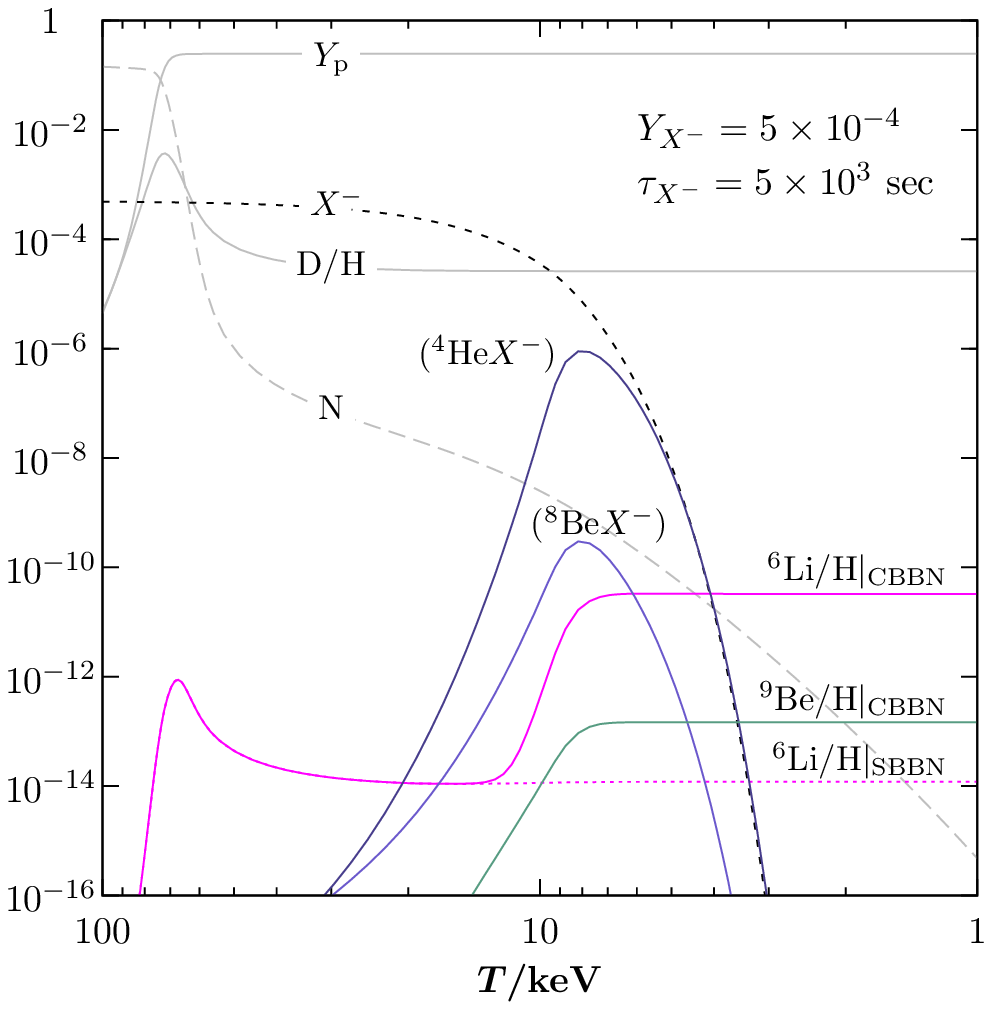}}
\caption{\base The temperature evolution of the bound state abundances
  (\hef\xm) and (\beet\xm), as well as the synthesis of \lisx\ and
  \ben\ at $T\sim$8 keV. The initial abundance of negatively charged
  particles is $Y_X = 5\times 10^{-4}$ and the lifetime is 5000~s.  }
\label{figure9}
\end{figure}

Current estimates and calculations of the CBBN rates are used to
determine the generic constraints on the lifetimes and abundances of
charged particles.  The essence of these limits is depicted in Figures
\ref{figure8} and \ref{figure9}, which shows that for typical \xm\
abundances the lifetime of the charged particles would have to be
limited by a few thousand seconds (unless a special mechanism for
suppressing $Y_X$ is found, as in {\it e.g.}
\cite{Ratz:2008qh,Pradler:2008qc}).  This is the most important fact
to be learned from the catalysis of BBN by charged particles. Although
to lowest order non-thermal BBN is sensitive to the energy density of
decaying particles, the CBBN processes are controlled by the number
density of \xm, which underscores the complementary character of these
constraints.  In some models, in which both catalysis and
non-equilibrium nucleosynthesis occur, catalysis dominates the
non-thermal production of \li6 for all particles with hadronic
branching ratio $B_h\simle 10^{-2}$~\cite{Jedamzik:2007qk}, whereas
\li7 destruction is usually dominated by neutron injection effects
unless $B_h\simle 10^{-4}$.

\subsubsection{ Catalysis below 1 keV?}

Finally we comment on the possibility of ($p$\xm) catalysis of nuclear
reactions, discussed in
Refs. \cite{Dimopoulos:1989hk,Jedamzik:2007cp}.  Although it is
conceivable that the absence of the Coulomb barrier for this compound
nucleus may lead to significant changes of SBBN/CBBN predictions, in
practice ($p$\xm)-related mechanisms are of only secondary importance
in most cases.  The large radius and shallow binding of this system
leads to a fast charge-exchange reaction on helium: ($p$\xm) + \hef
$\to$ (\hef\xm) + $p$. This reaction reduces the abundance of ($p$\xm)
below $10^{-6}$ relative to hydrogen, as long as $Y_{X^-}\simle Y_{\rm
  ^4He}$, making further reactions inconsequential for any observable
element \cite{Pospelov:2008ta}.  In the less likely case in which
$Y_{X^-}\simge Y_{\rm ^4He}$, significant late-time processing due to
($p$\xm) bound states may still occur. Nevertheless, such late-time
BBN typically leads to observationally unacceptable final BBN yields.

Unlike in the SBBN case and even in non-equilibrium nucleosynthesis
that utilizes mostly measured nuclear reaction rates, CBBN rates
cannot be measured in the laboratory; significant nuclear theory input
for the calculation of the reaction rates is required.  However,
because the $X^-$ participates only in electromagnetic interactions,
such calculations are feasible, and dedicated nuclear theory studies
\cite{Hamaguchi:2007mp,Kamimura:2008fx} for certain important CBBN
processes [such as those of Eqs.~(\ref{pburning}) and (\ref{cbbnLi6})]
have already been performed.

\subsection{Catalysis by strongly interacting relics}

Catalysis by strongly interacting relics, $X_s$, is another generic
possibility that has recently been addressed
\cite{Kusakabe:2009jt,Jonathan}. The main difficulty of this scenario
lies in determining the properties of the composite nuclei that
contain $X_s$.  For some types of $X_s$, some progress can be made.
Specifically, if $X_s$ represents an isospin doublet similar to
$(n,p)$, one could draw some conclusions about the binding energies
with nucleons by utilizing our extensive theoretical knowledge of
two-nucleon systems.  For example, a heavy scalar quark (or squark),
would attract one light quark and become a spin-1/2 ``mesino'' with
similar interaction properties as normal nucleons.  In other examples
involving a particle in the adjoint representation, such as a gluino,
it is almost impossible at this point to determine whether it would
bind to a single nucleon and, if so, with what energy. Regardless of
the type of strongly interacting relic, the relic's abundance is
expected to be suppressed by more than five orders of magnitude
compared with \xm\  due to a very efficient annihilation channel
following hadronization~\cite{Kang:2006yd}.

Detailed investigations of mesino-nucleon bound systems
\cite{Jonathan} offer insight into $X_s$-catalyzed BBN.  By making
fairly generous assumptions about the uncertainties in the interaction
strength, the binding energy of the mesino-nucleon system would
typically be between 5 and 40 MeV, which is much larger than the
deuteron binding energy~$E_d$.  A tighter bound nucleon-mesino system
is expected, because the shallow binding of the deuteron is to some
extent ``accidental'', and because going from a nucleon-nucleon system
to a nucleon-mesino decreases the kinetic energy and results in a
stronger binding. The main cosmological consequence of a tightly-bound
mesino-nucleon system is the formation of such bound states long {\it
  before} the end of the SBBN deuterium bottleneck, and, hence, at
times when neutrons are abundant. For example, a $\sim 20$~MeV binding
would correspond to a new bottleneck at $T\sim 1$~MeV, thereby opening
up an interesting new possibility for nucleosynthesis in the
neutron-rich environment {\it around} each $X_s$
particle. Unfortunately, a detailed study of such a nucleosynthetic
path is currently beyond theoretical capabilities, mostly due to the
rather uncertain spectrum of light mesino-containing nuclei. While
certain claims were made that nucleosynthesis proceeds to form the
$A\geq 6$ elements \cite{Kusakabe:2009jt}, it appears equally
plausible that nucleosynthesis around strongly interacting particles
stops at a ``compound helium''---a bound state consisting of a mesino
particle and three nucleons. Nucleosynthesis of heavier elements
around a mesino may be inhibited by $(p,\alpha)$ reactions, as well as
by an expected weak binding in a system containing five particles (one
mesino and four nucleons) \cite{Jonathan}. It is fair to say that much
more theoretical work will be required before we fully understand the
primordial catalysis caused by strongly-interacting particles.

\section{CONCLUSIONS}

Big Bang Nucleosynthesis, a very short period in the history of the
Universe, is an important reference point going back to cosmological
times of a few seconds.  In this review we have shown how new physics
can modify the synthesis of light elements and therefore be probed by
contrasting the observations of helium and deuterium (as well as
lithium, beryllium, and boron) with the BBN predictions.  All three
generic methods---modifications to the timing of main BBN events via
extra degrees of freedom or space-time dependent couplings and mass
scales, non-equilibrium nucleosynthesis triggered by energy injection
via the annihilation or decay of heavy particles, and particle
catalysis of BBN reactions---are very important to many extensions of
the Standard Model. In some models, the sensitivity of BBN to new
physics is exceptional and indeed exceeds all other direct and
indirect probes. A representative example is the non-trivial limit on
the lifetime of a metastable charged particle ({\it e.g.} a stau
decaying to a gravitino) that follows from the considerations of the
\lisx\ abundance, a limit that currently cannot be constrained other
than in BBN.

The amount of lithium predicted by SBBN is of the same order of
magnitude as the value indicated by the Spite plateau. However, a
detailed comparison with the standard cosmology and particle physics
input shows that SBBN over-predicts the abundance of \li7 by a factor
of three to five. This discrepancy has become more concrete, given
that the baryon-to-photon ratio and the main SBBN reactions
determining \lisv\ abundance are known to an accuracy of better than
10\%.  We have suggested possible solutions to the problem that may
reside in subtle astrophysical effects, or in new cosmological and/or
particle physics ingredients.  Here we summarize these (and some
additional) options

\begin{itemize}

\item The most economical solution to the lithium discrepancy would be
  an astrophysical mechanism that depletes lithium from the photic
  zone in the atmospheres of population II stars. Several ideas as to
  the nature of this mechanism is have been proposed in the
  literature, and hints at traces of such depletion in the scatter of
  points along the Spite plateau have been found
  \cite{Asplund:2005yt}.  Although some amount of depletion is
  certainly possible, it is premature to concede that it can account
  for the discrepancy of a factor of approximately three. Clearly that
  this issue remains one of the most important astrophysical issues
  in the cosmology of the early Universe.

\item Nuclear physics is unlikely to be responsible for the solution
  to the lithium problem, as all main reactions participating in
  creation and destruction of lithium are well known. Perhaps the last
  subject worth of detailed investigation is a possible resonant
  enhancement of the \bes+D reaction~\cite{Cyburt:2009cf}, which could
  contribute to the depletion of \bes, albeit at the very end of the
  most optimistic range for the parameters of such a resonance. This
  issue can be directly clarified through  experiments.

\item Given that lithium is observed in the Milky Way, whereas the
  measurements of deuterium and especially $\eta_b$ are global, one
  may speculate that baryons and possibly dark matter are distributed
  non-uniformly and that the 30-50\% downward fluctuation of $\eta_b$
  at the local patch of the Universe leads to a local value of lithium
  that is close to the Spite plateau value (see {\it e.g.}
  \cite{Holder:2009gd} for a recent discussion).  Of course, the
  variations in baryon number by a factor of a few, if persistent at
  all scales, would create very strong isocurvature-type features in
  the CMB polarization maps. However, it is also possible that models
  with anticorrelated baryon-dark matter fluctuations with a very blue
  perturbation spectrum may survive all observational constraints.

\item Particle physics models with unstable or annihilating particles
  may have an important effect on the lithium abundance. Although the
  photo-disintegration of lithium by a factor of three is difficult to
  reconcile with the deuterium abundance and/or the \het/D ratio
  (unless the mass of a decaying particle is carefully chosen), the
  energy injection with some yield of nucleons in the final state may
  fare much better.  Indeed, the injection of neutrons (regardless of
  the details of the particle physics mechanism) {\it before} the
  freeze-out of $\lisv(p,\alpha)\alpha$ reaction can reduce the \bes\
  and therefore \lisv\ abundance. A necessary consequence of this
  scenario is a somewhat elevated D/H abundance, and probably elevated
  concentrations of other rare isotope of lithium (and possibly
  beryllium and boron).

\item The catalytic suppression of \lisv\ by negatively charged
  particles relies on the fact that $(\bes X^-)$ is the deepest bound
  state among all \xm-bound nuclei which can form at temperatures of
  approximately 30 keV. The most efficient suppression of \bes\
  happens in models allowing for weak transitions to \lisv\ at these
  temperatures.  The need for a large number of \xm\ at 30~keV also
  implies the presence of \xm\ at $T\sim10$~keV, which would lead to
  the catalyzed synthesis of \lisx\ and \ben.

\item The last and perhaps most exotic option is the possibility of
  coupling constants and mass scales changing over time. Indeed, a
  $\sim 5$\% change in the deuterium binding energy can create a very
  strong downward shift of the \lisv\ abundance, while remaining
  consistent with helium and deuterium abundances.

\end{itemize}

We do not yet know how to resolve the lithium problem; only continuing
progress in particle physics, cosmology, and astrophysics will help to
clarify this intriguing discrepancy.

Finally, we discuss some anticipated future developments that are not
necessarily directly related to BBN but rather are related to
questions discussed in our review.

\begin{itemize}

\item CMB physics will continue to deliver very important results.  Of
  particular interest to this review is the increased sensitivity on
  small angular scales (high-l multipoles) where the effects of extra
  ``dark radiation'' states or certain types of isocurvature
  fluctuations are the most pronounced.  Future data from the Planck
  satellite experiment will allow us to probe the dark radiation
  energy density with an accuracy of $\Delta N_{\nu, \mathrm{eff}}
  \sim 0.5$ or better; results thus obtained will thereby provide an
  independent check of the constraints resulting from \hef\
  abundances.

\item The Large Hadron Collider is expected to deliver its first
  physics results in the near future. The TeV energy frontier will be
  effectively probed, which may lead to the discovery of stable or
  massive metastable particles. If such particles were neutral, their
  signature would constitute ``missing energy'', which is difficult
  but not impossible to disentangle from SM processes. However, if a
  metastable particle were charged a under strong or electromagnetic
  gauge group, the signatures would be much more spectacular, with
  long charged tracks that would allow us to identify the mass of such
  objects. If charged metastable states are discovered at the LHC, the
  theories of non-equilibrium and catalyzed nucleosynthesis would be
  further reinforced.

\item A massive expansion of the dark matter search program, both in
  space and in underground laboratories, may point to the existence of
  WIMPs in the near future.  If WIMPs turn out to be relatively light
  (in the 10-50 GeV window), then the energy injection in the early
  universe could lead to altered \lisv\ and \lisx\ abundances.  Heavier
  WIMPs with enhanced annihilation cross sections devised to fit the
  PAMELA positron anomaly~\cite{Adriani:2008zr} may also lead to
  changes in lithium abundances, and therefore any future development
  that supports or disfavors the WIMP interpretation of the PAMELA
  anomaly is of interest for the BBN predictions.

\item Finally, further theoretical progress in non-SBBN scenarios is
  expected.  Given the doubtful status of \lisx\ observations and its
  extreme fragility, it is highly desirable to perform calculations of
  beryllium and boron abundances generated in BBN scenarios with
  energy injection.  Doing so may place further constraints on
  annihilating/decaying WIMP scenarios. Additionally, more detailed
  calculations of some CBBN rates with a proper many-body nuclear
  physics input are needed.

\end{itemize}

\bibliographystyle{arnuke_revised}
\bibliography{Pospelov_Pradler_AnnRev}

\end{document}